\newif\ifmnras
\mnrasfalse
\ifmnras
	\documentclass[useAMS,usenatbib]{mn2e}
\else
	\documentclass[apj,numberedappendix]{emulateapj}
\fi
\usepackage{times}
\usepackage{graphics,epsf}
\usepackage{amsmath}                % American Mathematical Society package
\usepackage{amsfonts}               % American Mathematical Society fonts
\usepackage{amssymb}                % American Mathematical Society symbol
\usepackage{epsfig}                 % EPS figures
\usepackage{graphicx}
\usepackage{url}
\usepackage{tabularx}
\usepackage{booktabs}
\usepackage{subfigure}
\usepackage{graphicx}
\usepackage{grffile}
\def \cm{~\mathrm{cm}}
\def \s{~\mathrm{s}}

\def \km{~\mathrm{km}}
\def \kms{~\mathrm{km}~\mathrm{s}^{-1}}

\def \K{~\mathrm{K}}
\def \g{~\mathrm{g}}

\def \erg{~\mathrm{erg}}

\def \yr{~\mathrm{yr}}
\def \myr{~\mathrm{Myr}}

\def \kpc{~\mathrm{kpc}}

\usepackage{color}

\ifmnras
	\def \aap{A\&A}
	\def \araa{ARA\&A}
	\def \apj{ApJ}
	\def \apjl{ApJ}
	\def \apjs{ApJS}
	\def \nat{Nature}
	
	\def \mnras{MNRAS}

\fi

\ifmnras
	\title[Gentle heating by mixing]{Gentle heating by mixing in cooling flows}
	\author[S. Hillel \& N. Soker]{Shlomi Hillel and Noam Soker \\
	Department of Physics, Technion -- Israel, Institute of Technology, Haifa 32000, Israel;
	shlomihi@tx.technion.ac.il;
	soker@physics.technion.ac.il}
\fi

\begin{document}

\ifmnras
	\pagerange{\pageref{firstpage}--\pageref{lastpage}} \pubyear{2016}

	\maketitle
\else
	\title{Gentle heating  by mixing in cooling flow clusters}

	\author{Shlomi Hillel}
	\author{Noam Soker}
	\affil{Department of Physics, Technion -- Israel
	Institute of Technology, Haifa 32000, Israel;
	shlomihi@tx.technion.ac.il; soker@physics.technion.ac.il}
\fi

\label{firstpage}

\begin{abstract}
We analyze three-dimensional hydrodynamical simulations of the interaction of jets and the bubbles they inflate with the intra-cluster medium (ICM), and show that the heating of the ICM by mixing hot bubble gas with the ICM operates over tens of millions of years, and hence can smooth the sporadic activity of the jets. The inflation process of hot bubbles by propagating jets forms many vortices, and these vortices mix the hot bubble gas with the ICM. The mixing, hence the heating of the ICM, starts immediately after the jets are launched, but continues for tens of millions of years. We suggest that the smoothing of the active galactic nucleus (AGN) sporadic activity by the long-lived vortices accounts for the recent finding of a gentle energy coupling between AGN heating and the ICM.
\end{abstract}

% ==========================================================
\section{INTRODUCTION}
\label{sec:introduction}
% ==========================================================

The intra-cluster medium (ICM) in cooling flows, in galaxies, groups, and clusters of galaxies, is heated by jets launched from the central active galactic nucleus (AGN) and operate via a negative feedback mechanism (e.g., \citealt{Fabian2012, McNamaraNulsen2012, Farage2012, Gasparietal2013, Pfrommer2013, Baraietal2016}; for a recent review see \citealt{Soker2016}).
It is thought now that the feedback is closed by the \emph{cold feedback mechanism} \citep{PizzolatoSoker2005}, namely, cold dense clumps that feed the AGN
(e.g., some papers from the last 2 years, \citealt{Gaspari2015, VoitDonahue2015, Voitetal2015, Lietal2015, Prasadetal2015, SinghSharma2015, Tremblayetal2015, ValentiniBrighenti2015, ChoudhurySharma2016, Hameretal2016, Loubseretal2016, Russelletal2016, McNamaraetal2016, YangReynolds2016b, Baraietal2016, Prasadetal2016, Tremblayetal2016, Donahueetal2017, GaspariSadowski2017, Gasparietal2017, Voitetal2017, Meeceetal2017}).
The new results of \cite{Hoganetal2017} suggest that the perturbations that feed the AGN should start as non-linear ones, as was suggested in the original paper by \cite{PizzolatoSoker2005}.

Although there is a general consensus on the AGN feedback activity, there is a dispute on the exact process that transfers the energy from the jets to the ICM in this jet feedback mechanism (JFM). Heating processes that have been proposed in the literature include sound waves (e.g., \citealt{Fabianetal2006, Fabian2012, Fabianetal2017}) that can be excited by jet-inflated bubbles \citep{SternbergSoker2009}, shocks that are excited by the jets (e.g., \citealt{Formanetal2007, Randalletal2015}; for problems with shock heating see, e.g.,  \citealt{Sokeretal2016}), heating by dissipation of ICM turbulence (e.g., \citealt{DeYoung2010, Gasparietal2014, Zhuravlevaetal2014}; for problems and limitations of turbulent heating see, e.g., \citealt{Falcetaetal2010, Reynoldsetal2015, Hitomi2016, HillelSoker2017}), cosmic rays (e.g., \citealt{Fujitaetal2013, FujitaOhira2013}), and mixing of hot bubble gas with the ICM (e.g., \citealt{BruggenKaiser2002,  Bruggenetal2009, GilkisSoker2012,HillelSoker2014, HillelSoker2016, YangReynolds2016b}).
Some processes can operate together, such as cosmic rays and thermal conduction (e.g., \citealt{GuoOh2008}), mixing of cosmic rays from jet-inflated bubbles to the ICM \citep{Pfrommer2013}, and heating by turbulence and turbulent-mixing (e.g. \citealt{BanerjeeSharma2014}).

The heating by mixing process is caused by the many vortices that are excited by the inflation process of the bubbles \citep{GilkisSoker2012,HillelSoker2014, HillelSoker2016, YangReynolds2016b}. A by product of this process is that the vortices induce turbulence in the ICM, accounting for the finding of turbulence in some cooling flows (e.g., \citealt{Zhuravlevaetal2014, Zhuravlevaetal2015, Arevalo2016, AndersonSunyaev2016, Hofmannetal2016}). {{{{ Although the vortices are very efficient in mixing hot bubble's gas with the ICM, the mixing does not necessary destroy the bubbles. In two dimensional (2D) simulations we have found that bubbles continue to rise as they induce vortices in their surroundings \citep{HillelSoker2014}. The heating by mixing can account for both turbulence in the ICM and for the presence of bubbles at large distances from the center.  }}}}

The JFM is more complicated than what simple arguments might suggest. For example, the jets might have positive components to the feedback cycle in addition to the more influential negative one.
The positive components include the interaction of the jets with the ICM that form inhomogeneities that are the seeds of future dense blobs (e.g., \citealt{PizzolatoSoker2005}), or the bubbles that through their buoyant motion lift low entropy gas that can cool and fall to feed the AGN (e.g., \citealt{McNamaraetal2016}).

In the present study we refer to two recent papers, and we further emphasize the dominant role that vortices that are formed during the inflation process of the bubbles play in the feedback  heating cycle of the ICM.

\cite{SternbergSoker2008} show that to obtain the correct flow structure it is mandatory to inflate bubbles by jets, rather than by artificially injecting energy off-center. In a recent paper \cite{Weinbergeretal2017} insert jets off-center. One of their conclusions is that mixing of lobe material with the ICM is sub-dominant in the heating process.
In section \ref{sec:inflation} we examine the formation of vortices early on by jets injected from the center. We argue that to obtain the full power of heating by mixing, the jets should be inserted from the center.

In a new and thorough study \cite{Hoganetal2017} analyze the properties of 56 clusters of galaxies, and conclude that ``. . .the energy coupling between AGN heating and atmospheric gas is gentler than most models predict'' (see also \citealt{McNamaraetal2016}).
In an earlier paper \citep{HillelSoker2016} we conducted three dimensional (3D) hydrodynamical simulations of intermittent jets interacting with the ICM. Each activity phase lasts for a period of $10 \myr$, with a quiescent period of $10 \myr$ between two consecutive active phases. An interesting finding of these 3D hydrodynamical numerical simulations is that the large scale vortices continue to exist even in the quiescent periods.
This implies that the mixing is a continuous process, and no large variations are expected during the evolution if the decay time of the vortices is about equal or larger than the quiescent phases period.
In section \ref{sec:gentle} we show that the expectation of the heating by mixing process and the new findings of \cite{Hoganetal2017} are compatible with each other. {{{{ In a paper posted very recently, \cite{Zhuravlevaetal2017} argue that turbulent heating processes support a model of gentle AGN feedback. But as we commented before \citep{HillelSoker2017}, we think that mixing-heating is more efficient than turbulent heating. }}}}

We summarize our claims in section \ref{sec:summary}. We will present results from our earlier simulations, but the analysis extends to a new domain. We open by describing our numerical scheme in section \ref{sec:numerics}.

 % ==========================================================
\section{NUMERICAL SETUP}
\label{sec:numerics}
% ==========================================================

We present results from our earlier 3D hydrodynamical numerical simulations \citep{HillelSoker2016}, where we used the numerical code {\sc pluto} \citep{Mignone2007}. We further analyzed these simulations in our study of the galaxy group NGC~5813 \citep{Sokeretal2016}, and in our interpretation of the Hitomi observations of the Perseus cluster of galaxies \citep{HillelSoker2017}. We here describe only the essential features of the numerical scheme

The computational grid is in the octant where the three coordinates are
positive $0 \le x \le 50 \kpc$, $0 \le y \le 50 \kpc$ and $0 \le z \le 50 \kpc$ , and the $z$ axis is chosen along the symmetry axis of the jet. The $z=0$ plane is a symmetry plane, {{{{ as in reality two opposite jets are launched simultaneously, while here we simulate only one jet. At the inner planes, $x = 0$, $y = 0$ and $z = 0$, we apply reflective boundary conditions. On the outer boundaries, $x = 50 \kpc$, $y = 50 \kpc$ and $z = 50 \kpc$, we apply outflow boundary conditions. Heat conduction and viscosity are not included in the simulations. }}}}
The highest resolution of the adaptive mesh refinement is $\approx 0.1 \kpc$.

We inject the jet from a circle $\sqrt{x^2 + y^2} \leq 3 \kpc$  at the plane $z = 0$, and with a half-opening angle of $\theta_{\rm j} = 70^\circ$. The initial jet velocity is $v_{\rm j} = 8200 \kms$. The jet is periodic in time. It is injected continuously for a period of $10 \myr$, starting at $t=0$, followed by an off-phase that lasts for $10 \myr$. Namely, the jet-active phases are in the time intervals
\begin{equation}
20(n-1) \le t_n^{\rm jet} \le 10 (2n-1), \qquad n=1,2,3 \dots .
\label{eq:jet}
\end{equation}
The mass deposition rate into the two opposite jets (only one is simulated, or more accurately, only quarter of a jet is simulated) and the power of the two jets during each on-episode are $ \dot{M}_{2{\rm j}} = {2 \dot E_{2{\rm j}}}/{v_{\rm j}^2} = 94 M_{\odot}~\yr^{-1}$ and $\dot E_{2{\rm j}} = 2 \times 10^{45} \erg \s^{-1}$, respectively.
{{{{ The density of the jet at injection is about $10^{-26} \g \cm^{-3}$, much below the ambient density. The initial temperature of the jet is equal to that of the ambient medium $3\times 10^7 \K$. Therefore, at injection, the pressure of the jet is much below that of the ambient medium, and its initial thermal energy is negligible with respect to its initial kinetic energy (about 3 per cent).   }}}}
Such massive sub-relativistic wide outflows are supported by observations (e.g., \citealt{Aravetal2013}).

The initial density of the ICM in the grid is set to be (e.g., \citealt{VernaleoReynolds2006})
\begin{equation}
\rho_{\rm ICM}(r) = \frac{\rho_0}{\left[ 1 + \left( r / a \right)
^ 2 \right] ^ {3 / 4}},
\label{eq:den}
\end{equation}
with $a = 100 \kpc$ and $\rho_0 = 10^{-25} \g \cm^{-3}$. The initial ICM temperature is $T_{\rm ICM} (0) = 3 \times 10^7 \K$.
We include a gravity field that maintains an initial hydrostatic equilibrium, and we keep it constant in time. We also include radiative cooling for a solar metallicty gas from Table 6 of \cite{SutherlandDopita1993}.

% ==========================================================
\section{VORTICES BY JET-INFLATED BUBBLES}
 \label{sec:inflation}
% ==========================================================

\begin{figure} % [!htb]
\centering
\subfigure{\includegraphics[width=0.24\textwidth]{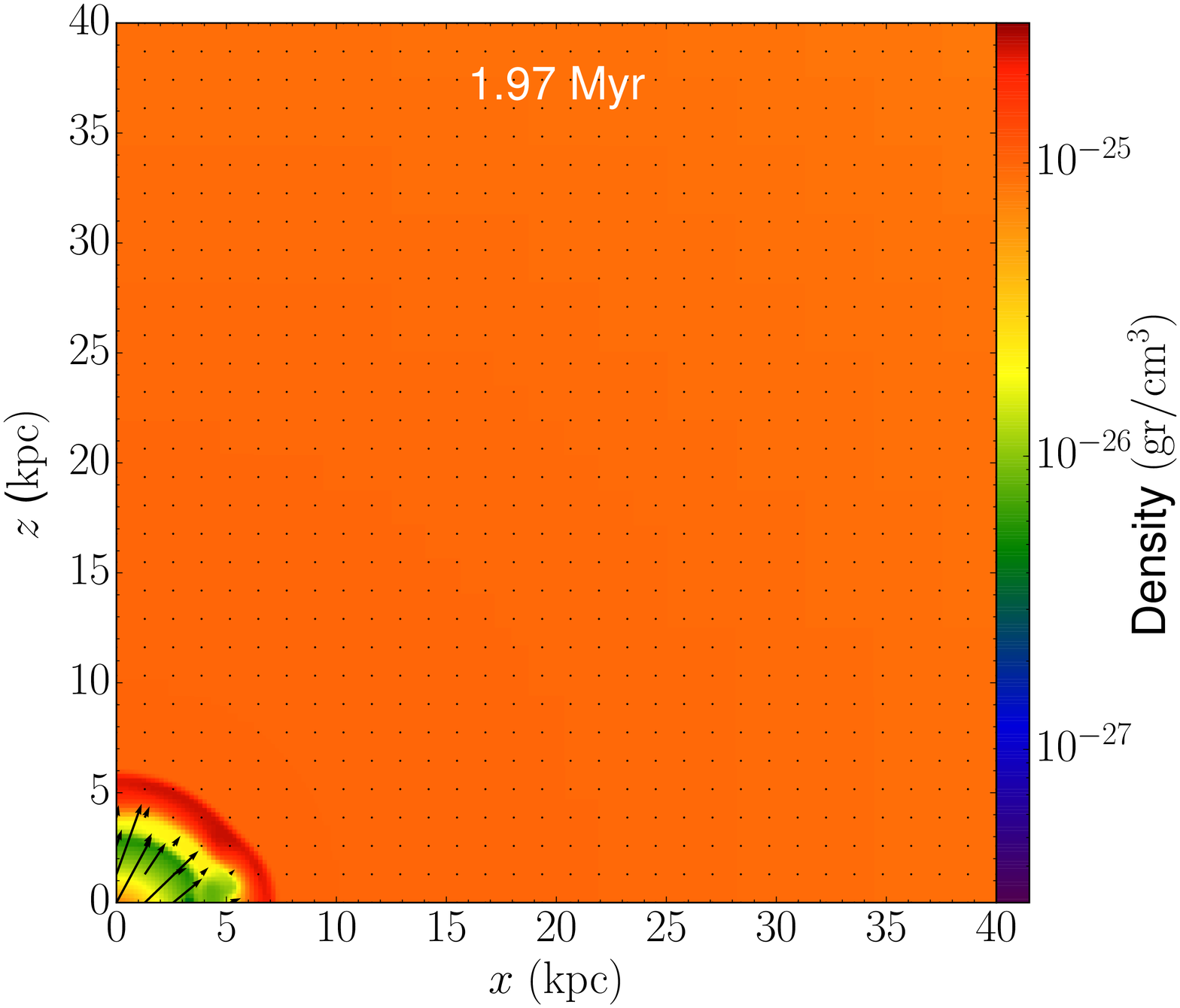}}
\hskip -0.25 cm
\subfigure{\includegraphics[width=0.24\textwidth]{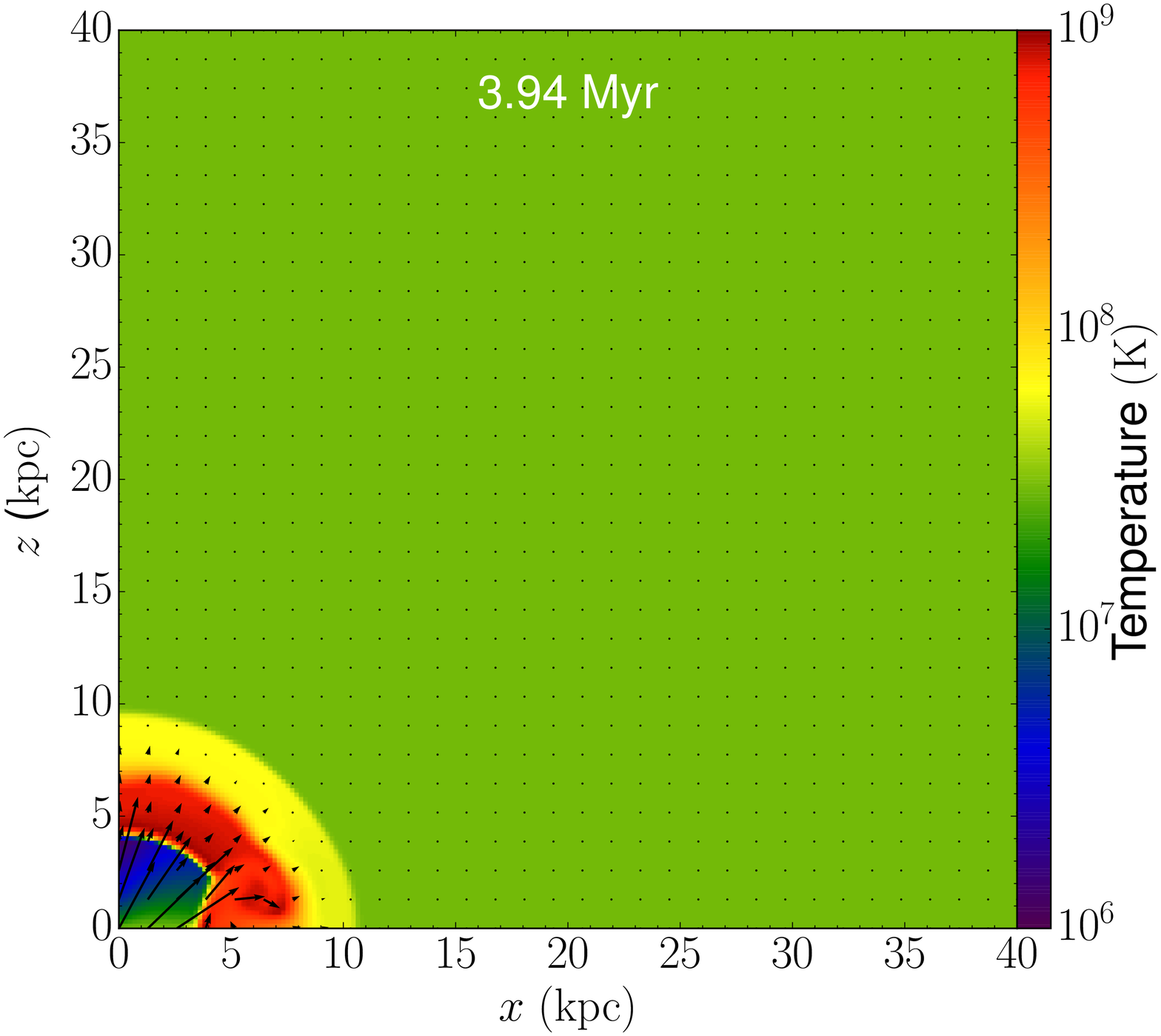}}
\subfigure{\includegraphics[width=0.24\textwidth]{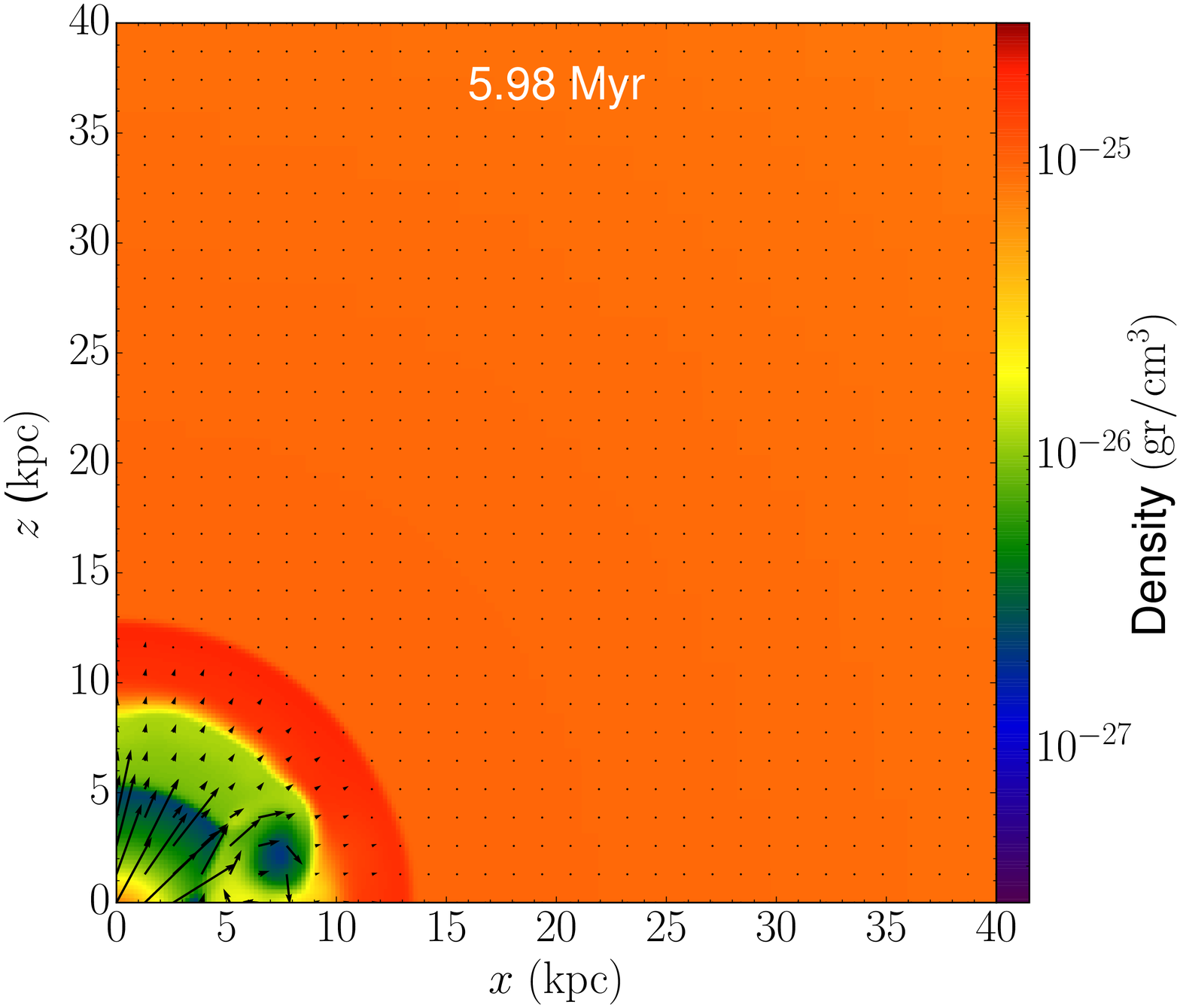}}
\hskip -0.25 cm
\subfigure{\includegraphics[width=0.24\textwidth]{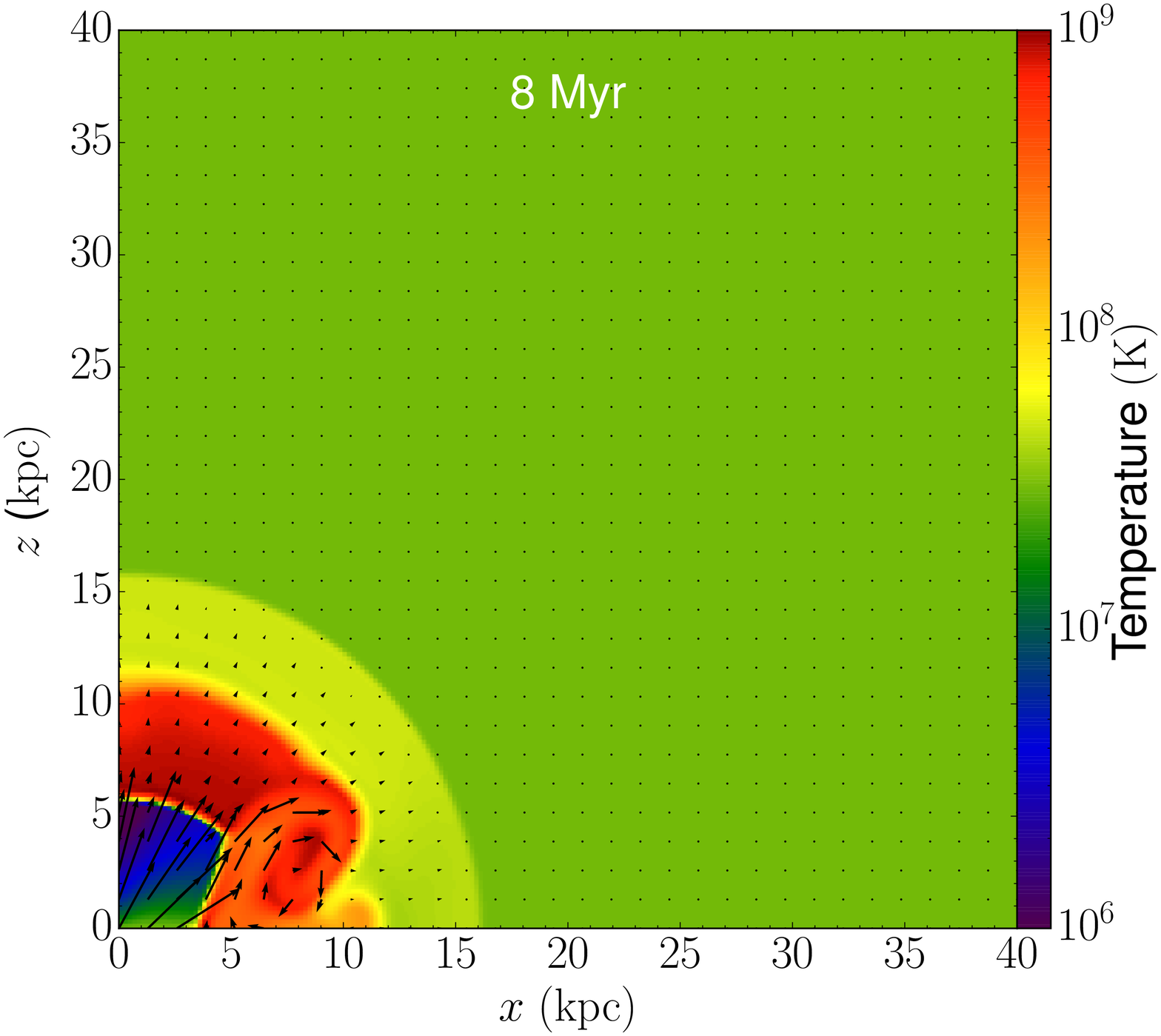}}
\subfigure{\includegraphics[width=0.24\textwidth]{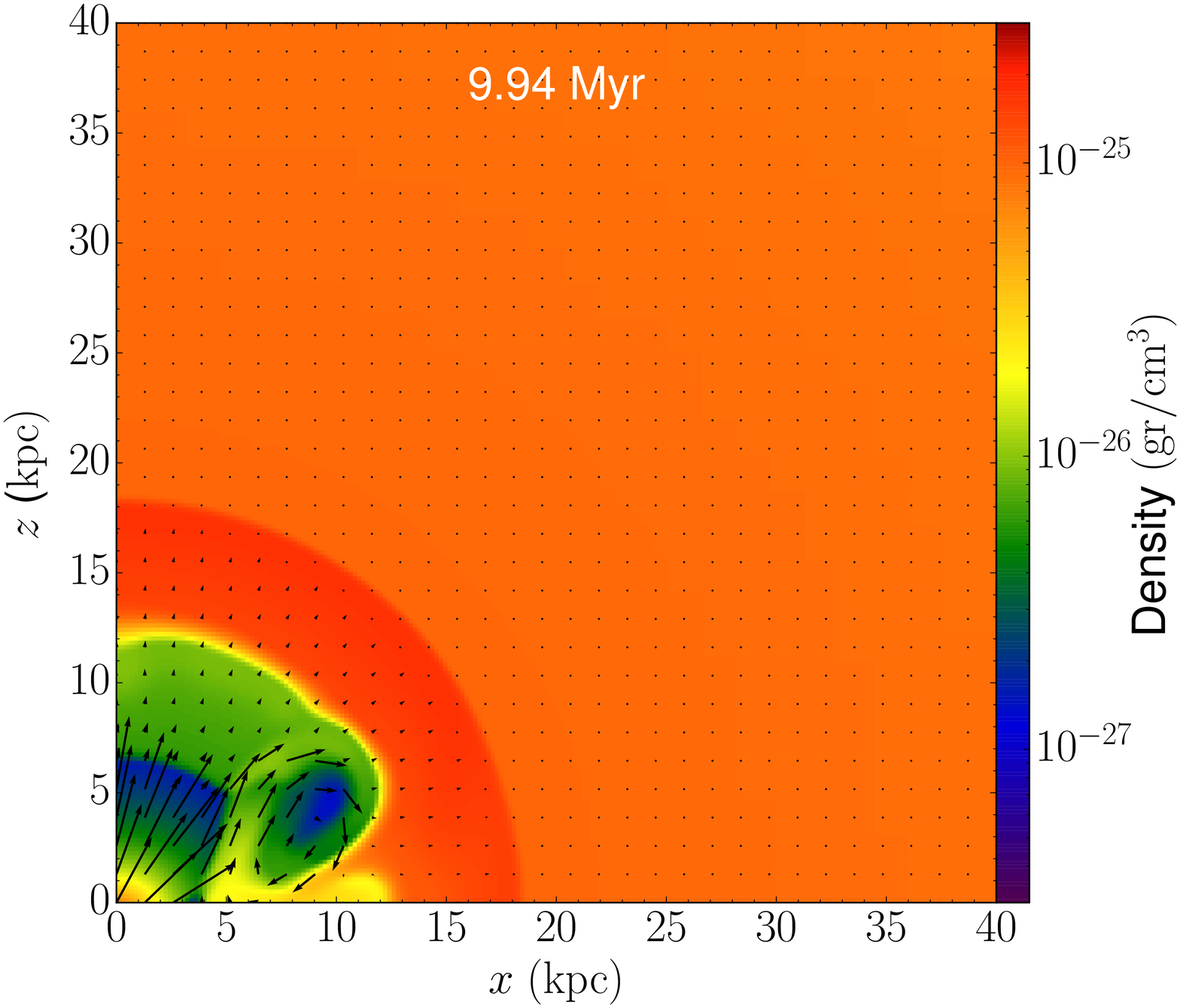}}
\hskip -0.25 cm
\subfigure{\includegraphics[width=0.24\textwidth]{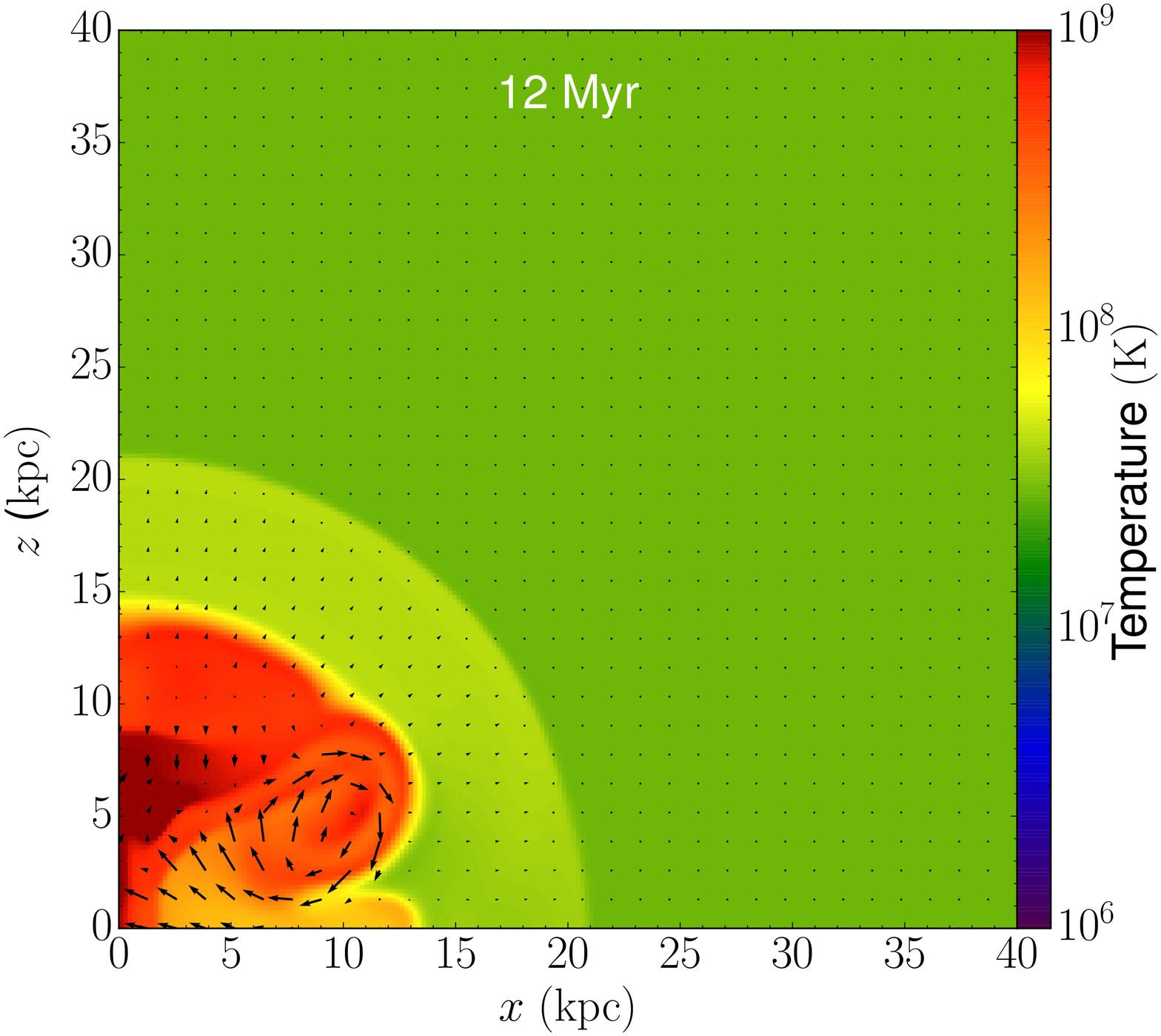}}
\subfigure{\includegraphics[width=0.24\textwidth]{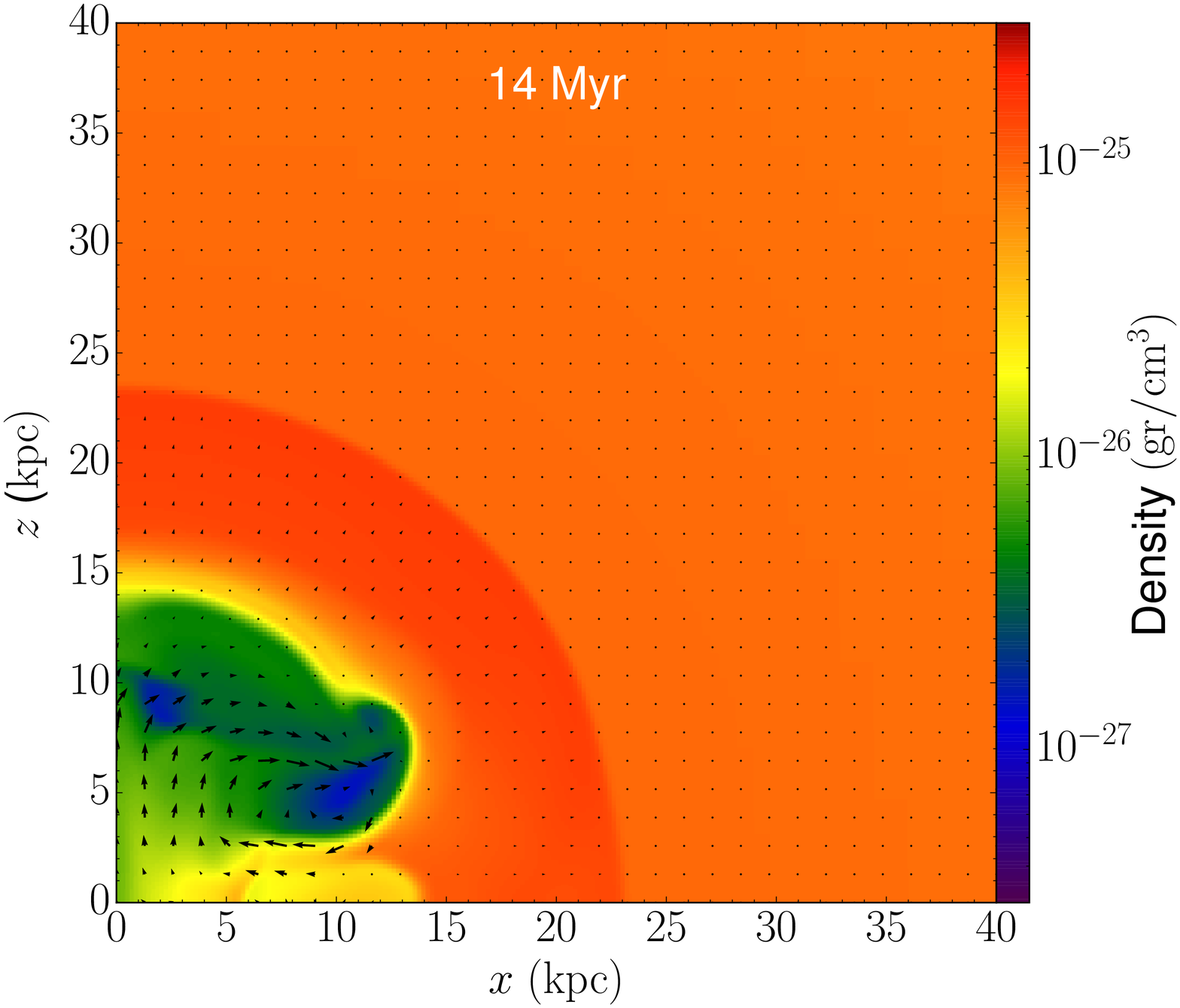}}
\hskip -0.25 cm
\subfigure{\includegraphics[width=0.24\textwidth]{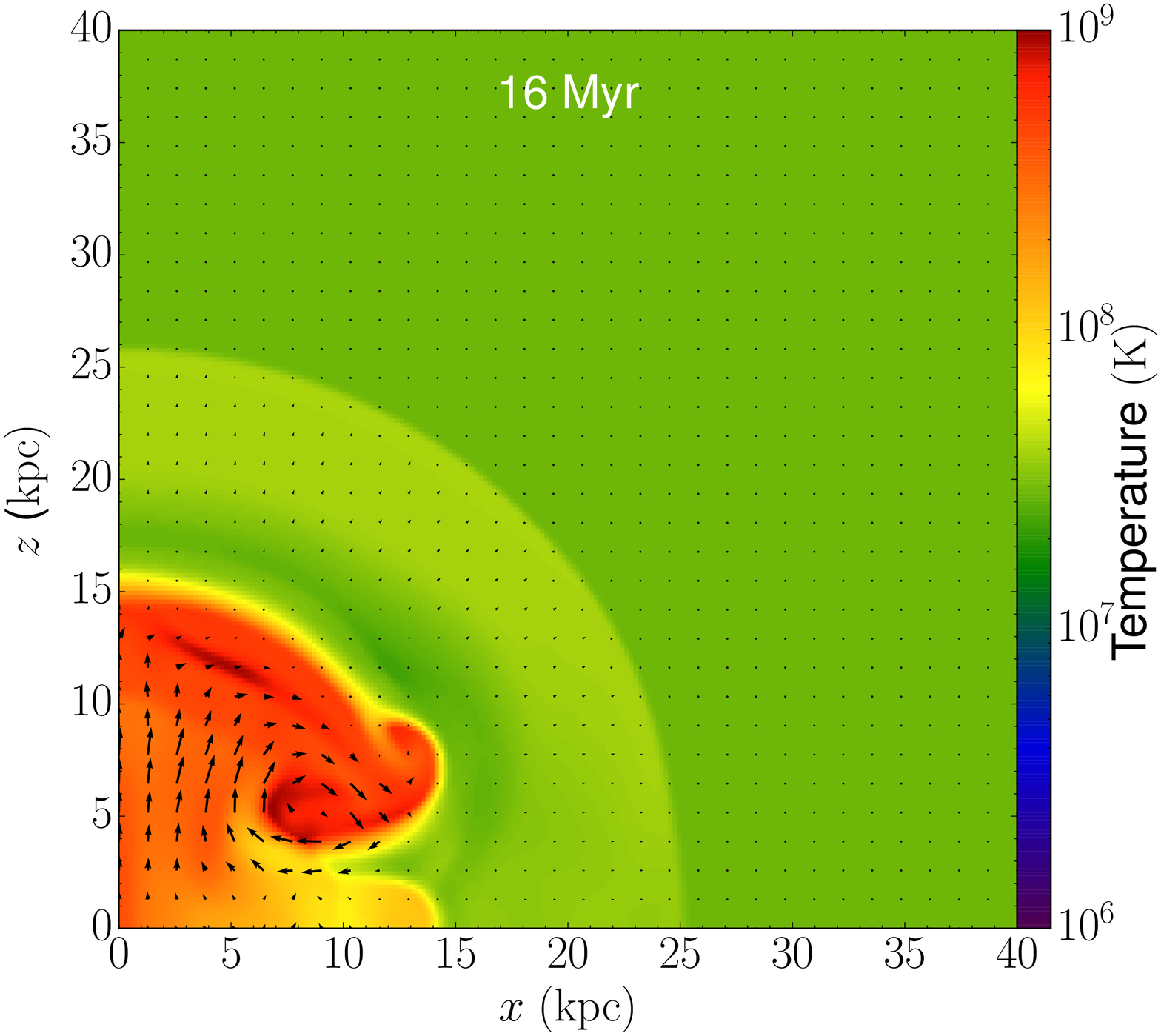}}
\subfigure{\includegraphics[width=0.24\textwidth]{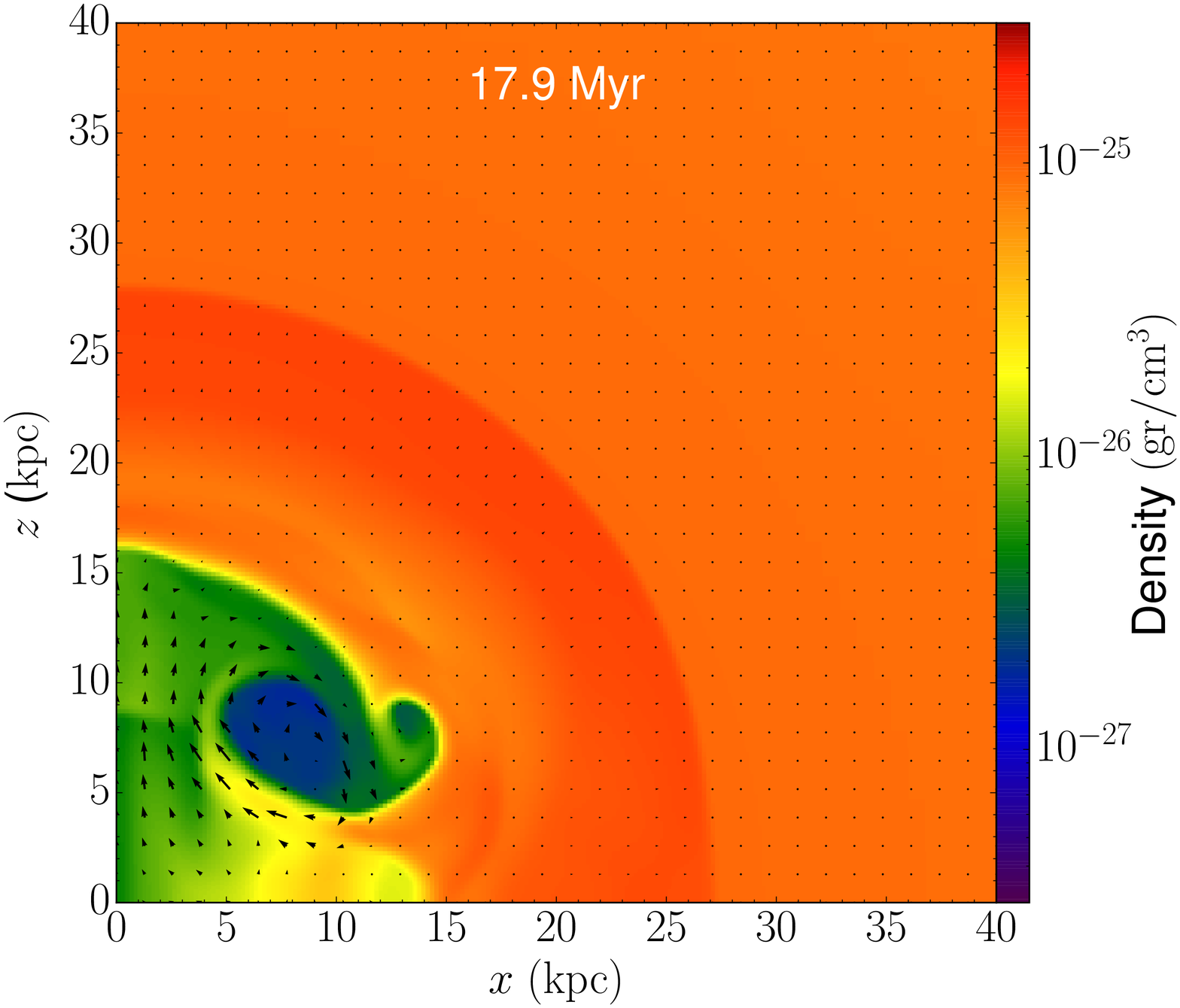}}
\hskip -0.25 cm
\subfigure{\includegraphics[width=0.24\textwidth]{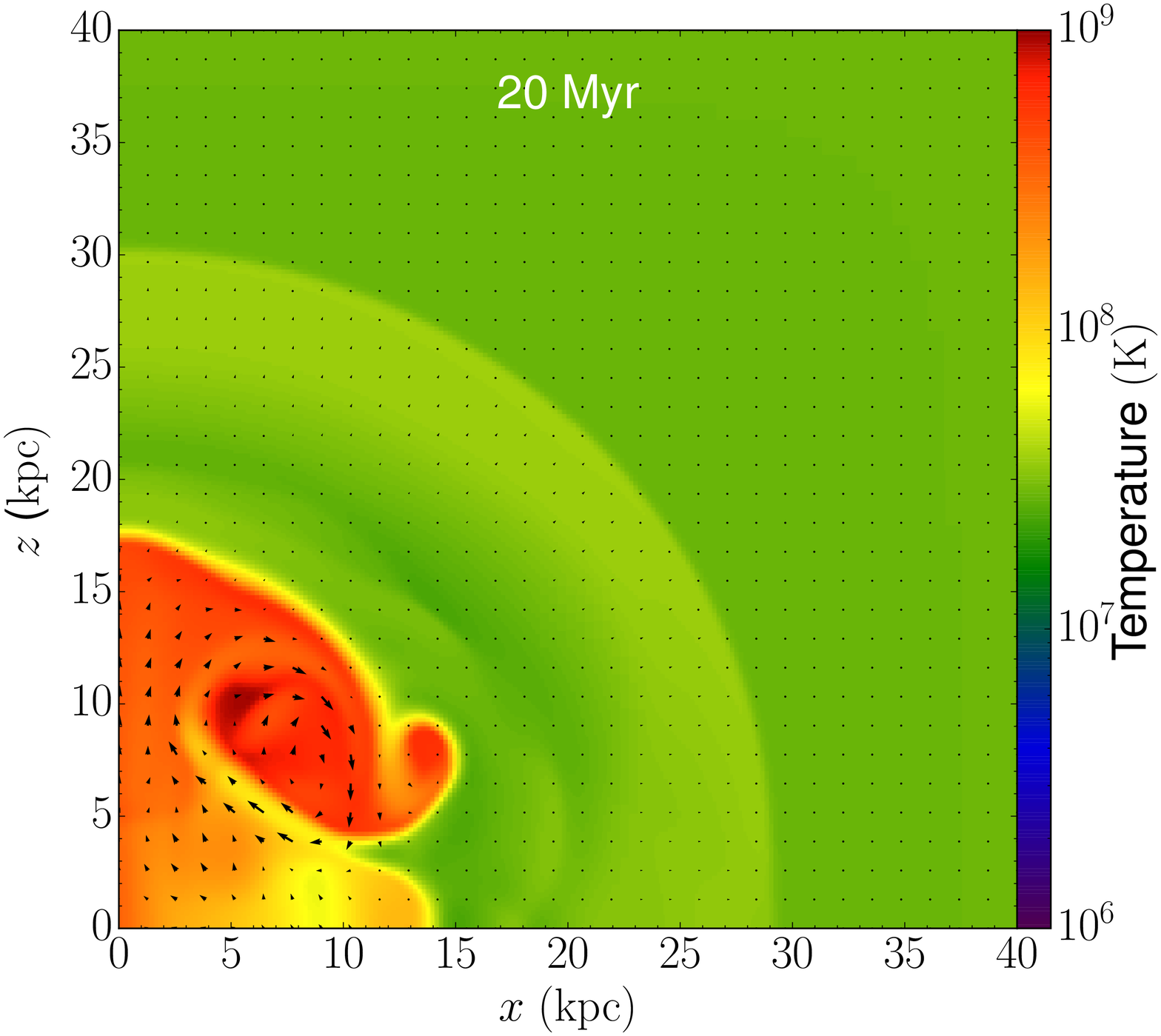}}
\caption{Early evolution of the flow velocity and either temperature or density,  presented in the $y=0$ meridional plane at the first jet's activity cycle. The color scales of the temperature and the density are logarithmic. Arrows show the velocity, with length proportional to the velocity magnitude. A length of $1 \kpc$ on the map corresponds to $1700 \km \s^{-1}$. When the jet is active, the length of arrows close to the origin corresponds to $8200 \kms$. }
\label{fig:EarlyFlow}
\end{figure}

In Fig. \ref{fig:EarlyFlow} we present the inflation of the bubble in the first activity cycle: the jet is active in the time period $0-10 \myr$ and it is turned off for the time period $10-20 \myr$. At $t=20 \myr$ the second activity cycle starts (eq. \ref{eq:jet}). We present the density (left column) and temperature (right column) in color-contours, and the velocity vectors by arrows, in the meridional plane $y=0$ at 10 times as indicated in the figure.
%FFFFFFFFFFFFFFFFFFFFFFFFFFFFFFFFFFFFFFFFFFFFFFFFFFF
%FFFFFFFFFFFFFFFFFFFFFFFFFFFFFFFFFFFFFFFFFFFFFFFFFFF

Fig. \ref{fig:EarlyFlow} clearly shows the rapid development of vortices inside and outside the jet-inflated hot bubble. The rapid development of vortices can be understood as follows. Although a large shear exists between the fast jet and the static ICM, the vortices are mainly formed in the post-shock region. As the jet's material hits the ICM it is shocked. A very high pressure region is formed in the post-shock region. The distance from the center of this high-pressure region increases with time as the jet continues to be active. The post-shock gas expands rapidly to the sides of that region, i.e., about perpendicular to the original direction of the jet. It then expands backward, to form what is termed the cocoon, i.e., shocked jet's material that lags behind the jet. This motion to the side and then backward forms large vortices.
The outward motion of the high pressure region and the vortex that it forms can be best seen by following the low-density region (that is a cross section of a low density volume) at the center of the vortex on the plane of the image (the $y=0$ plane). This low-density region is the blue-color region moving from about $(x,z)=(7,2) \kpc$ at $t=6 \myr$ to about $(x,z)=(9,7) \kpc$ at $t=18 \myr$ .

The lower right panel of Fig. \ref{fig:EarlyFlow} shows that even $10 \myr$ after the jets has been turned off ($t=20 \myr$) the temperature is not smoothed yet. This implies that the vortices did not completely mix yet the hot bubble gas and the ICM. We see also that the vortices, in particular the large vortex,  still exist and they continue the mixing process. We discuss this in detail in section \ref{sec:gentle} below.

The simulation of a propagating jet is essential to capture the formation of the vortices. Jets that encounter an ambient gas excite vortices even when the medium is homogenous, rather than stratified. When an artificial bubble or jet is inserted at zero velocity off-center, on the other hand, vortices might be formed only as a result of the upward motion of the hot region due to buoyancy. In an homogeneous medium artificial bubbles will form no vortices.
Other limitations of artificially introduced jets and bubbles are discussed by \cite{SternbergSoker2008}.

Let us summarize and further emphasize the analysis of our simulations in this section.
The many vortices that are formed by the shocked jets and the bubble-inflation process play a crucial role in heating the ICM and in determining the properties of the feedback cycle, not only in cooling flows, but in other environments as well \citep{Sokeretal2013, Soker2016}.
To obtain these vortices the numerical simulations must include propagating jets that start from the center. Simulations that inject static hot gas or jets off-center might lead to inaccurate conclusions. In a recent paper \cite{Weinbergeretal2017} insert jets off-center. We think that this is the reason that they find that mixing of material from the bubbles with the ICM is not the main heating mechanism of the ICM. When we inject jets from the center, we find heating by mixing to be the main heating mechanism of the ICM \citep{HillelSoker2016}.

% ==========================================================
\section{LONG LIVED VORTICES}
\label{sec:gentle}
% ==========================================================

\begin{figure} % [!htb]
\centering
\centering
\subfigure{\includegraphics[width=0.24\textwidth]{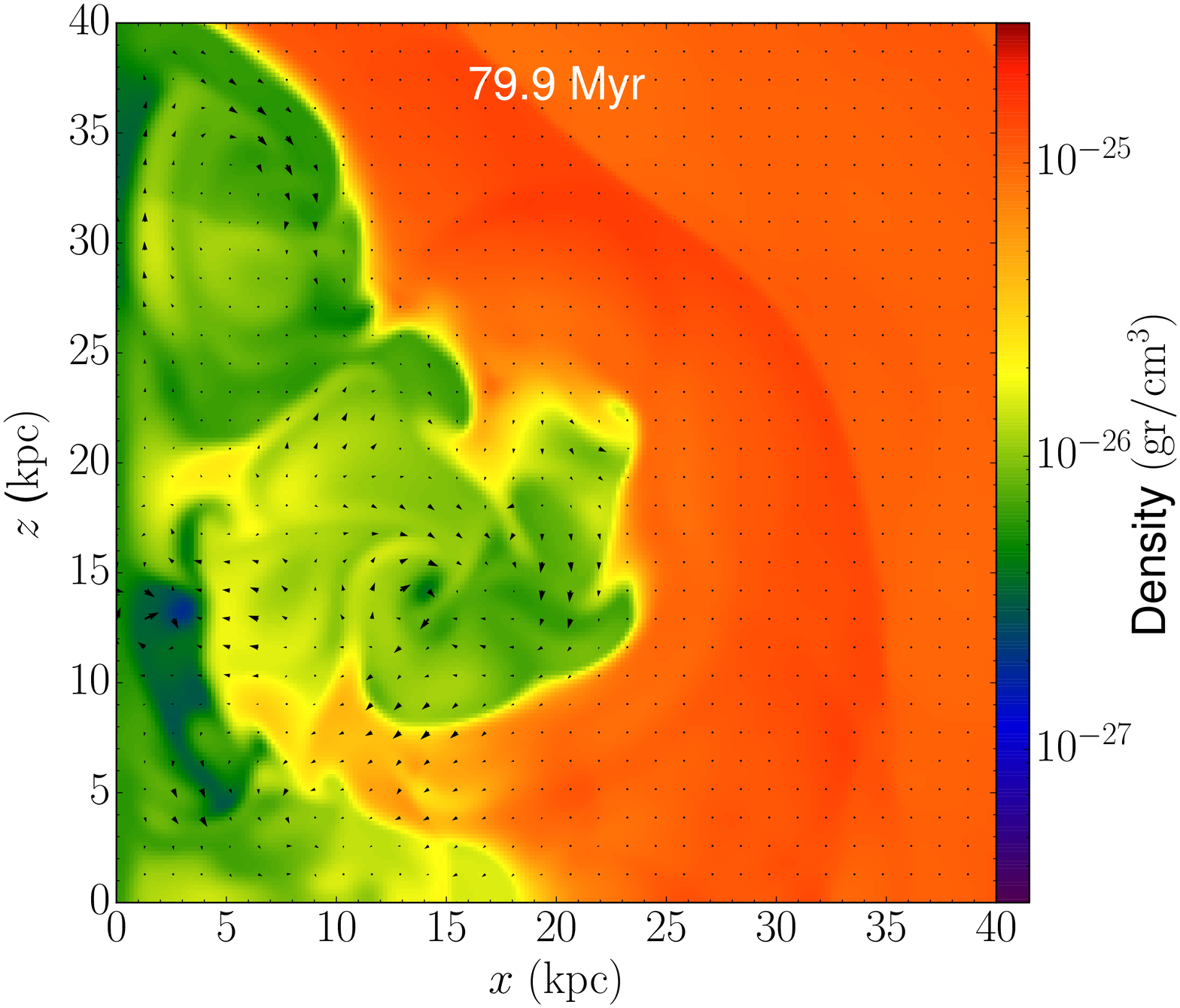}}
\hskip -0.25 cm
\subfigure{\includegraphics[width=0.24\textwidth]{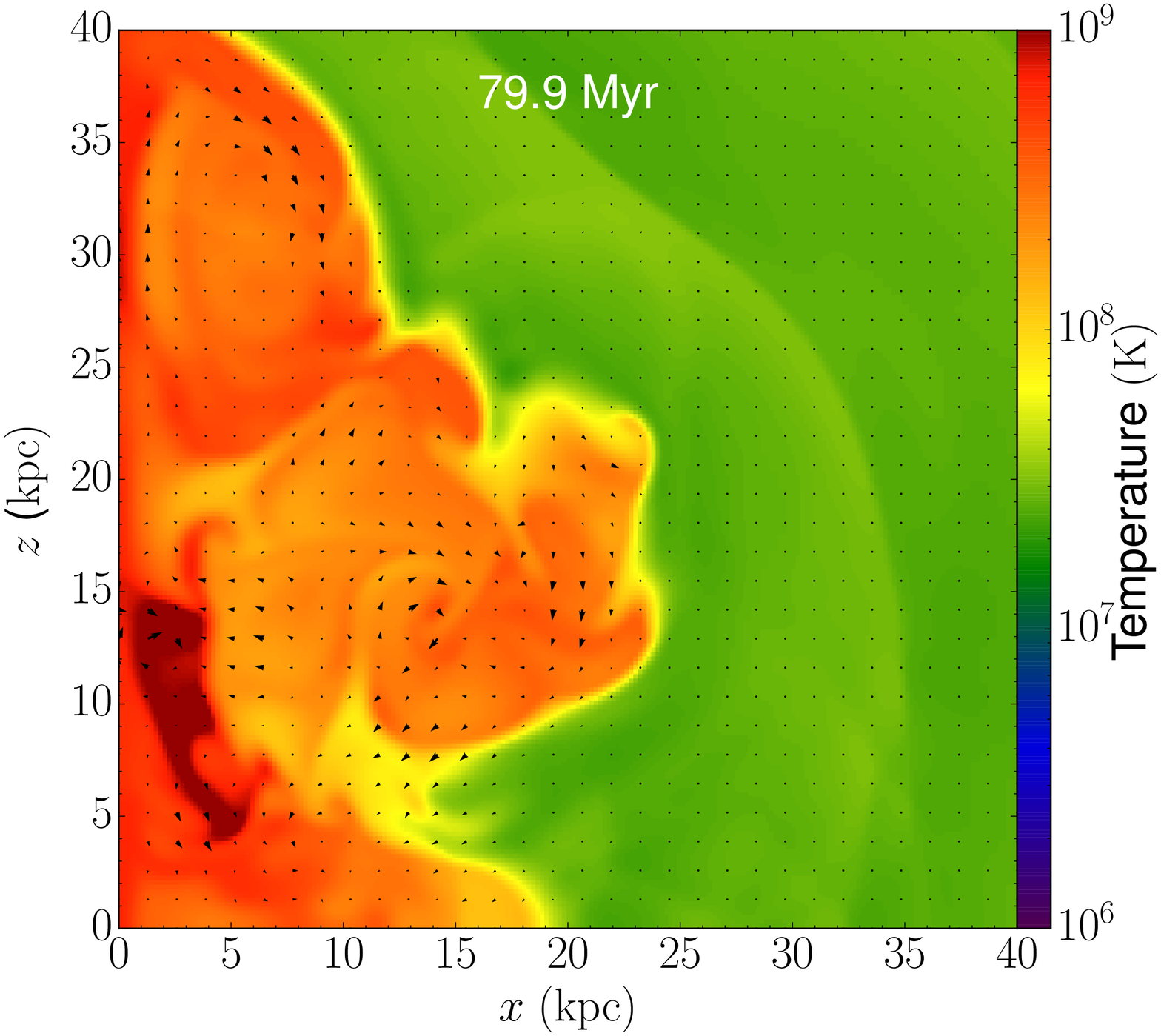}}
\subfigure{\includegraphics[width=0.24\textwidth]{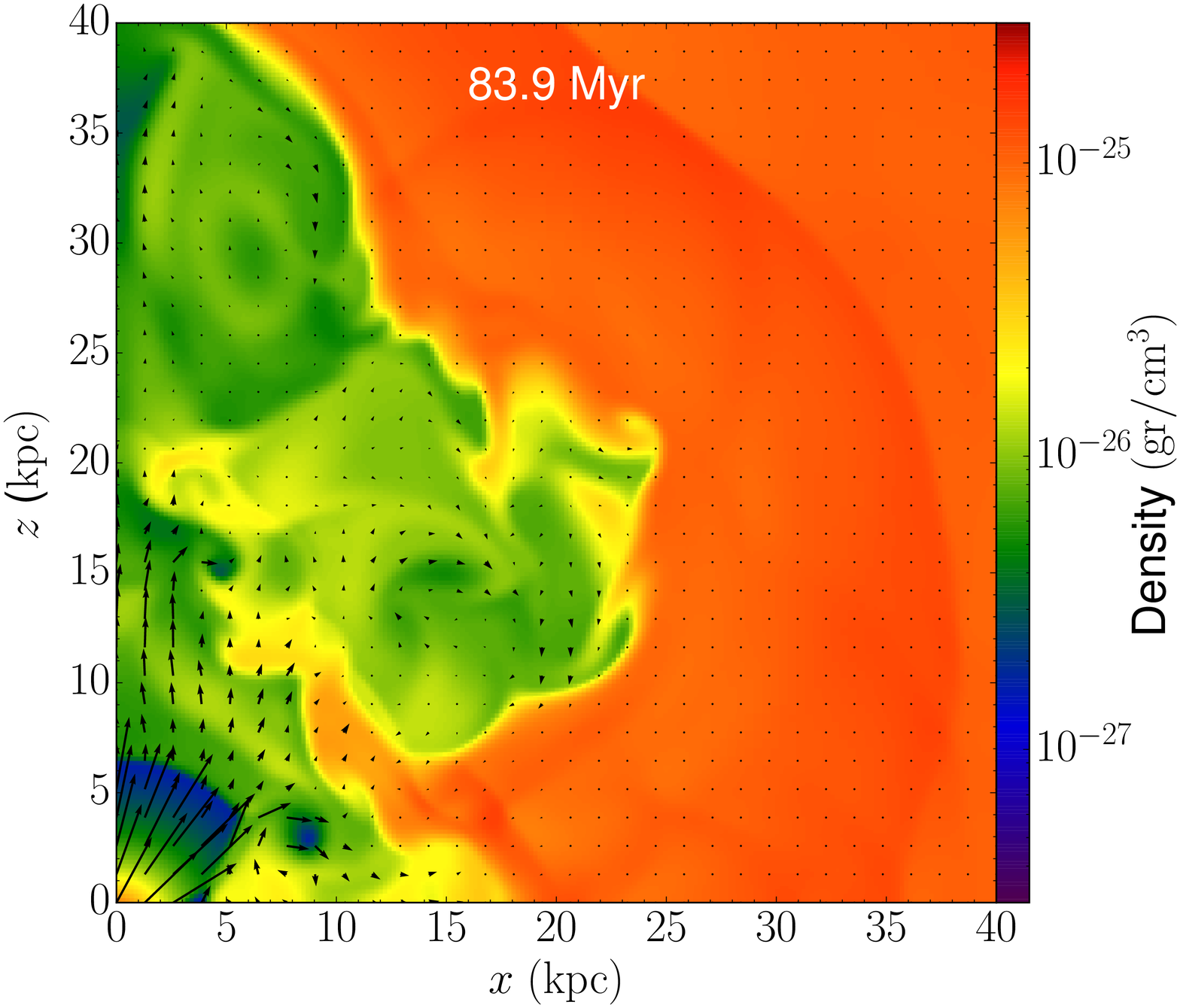}}
\hskip -0.25 cm
\subfigure{\includegraphics[width=0.24\textwidth]{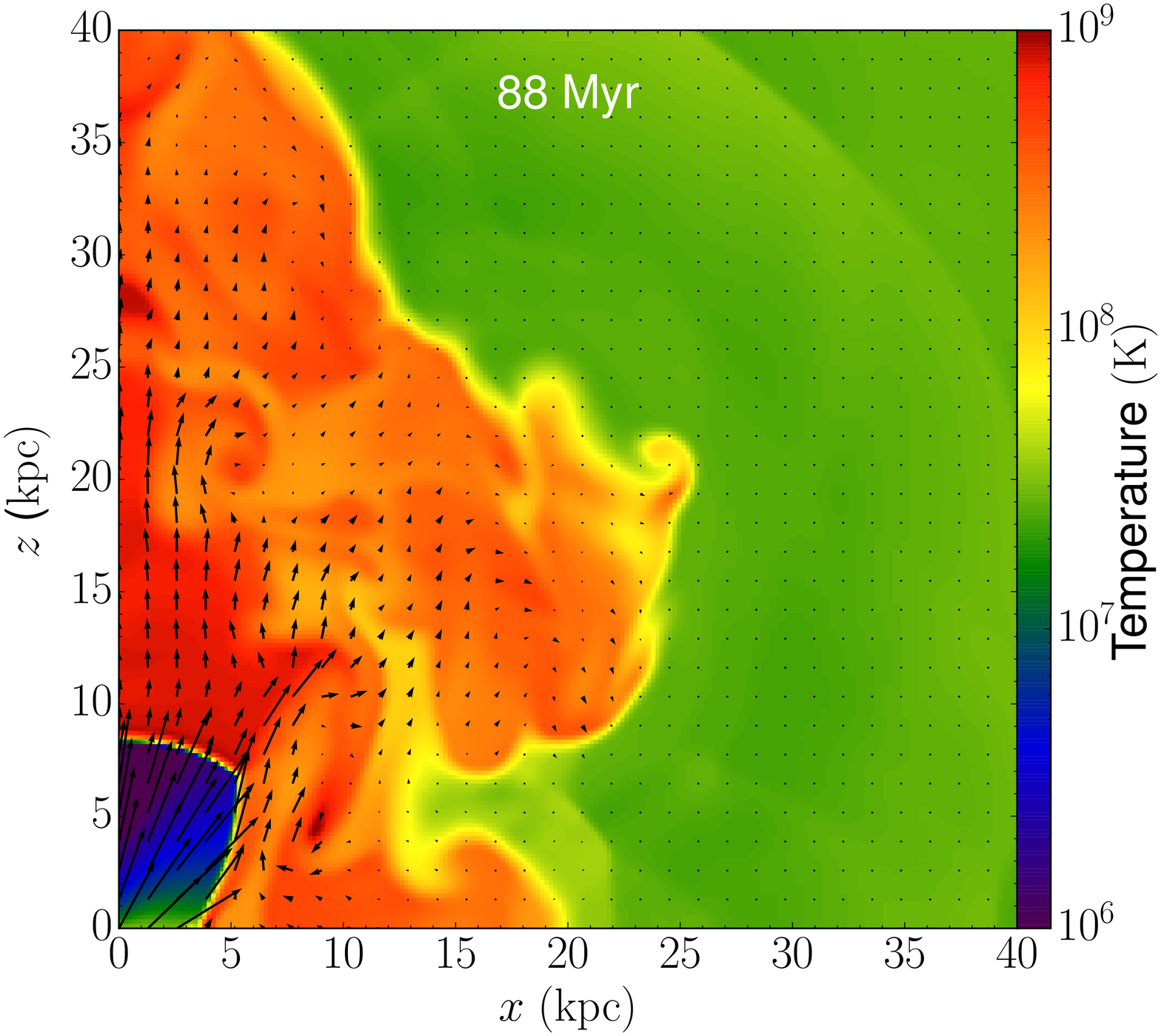}}
\subfigure{\includegraphics[width=0.24\textwidth]{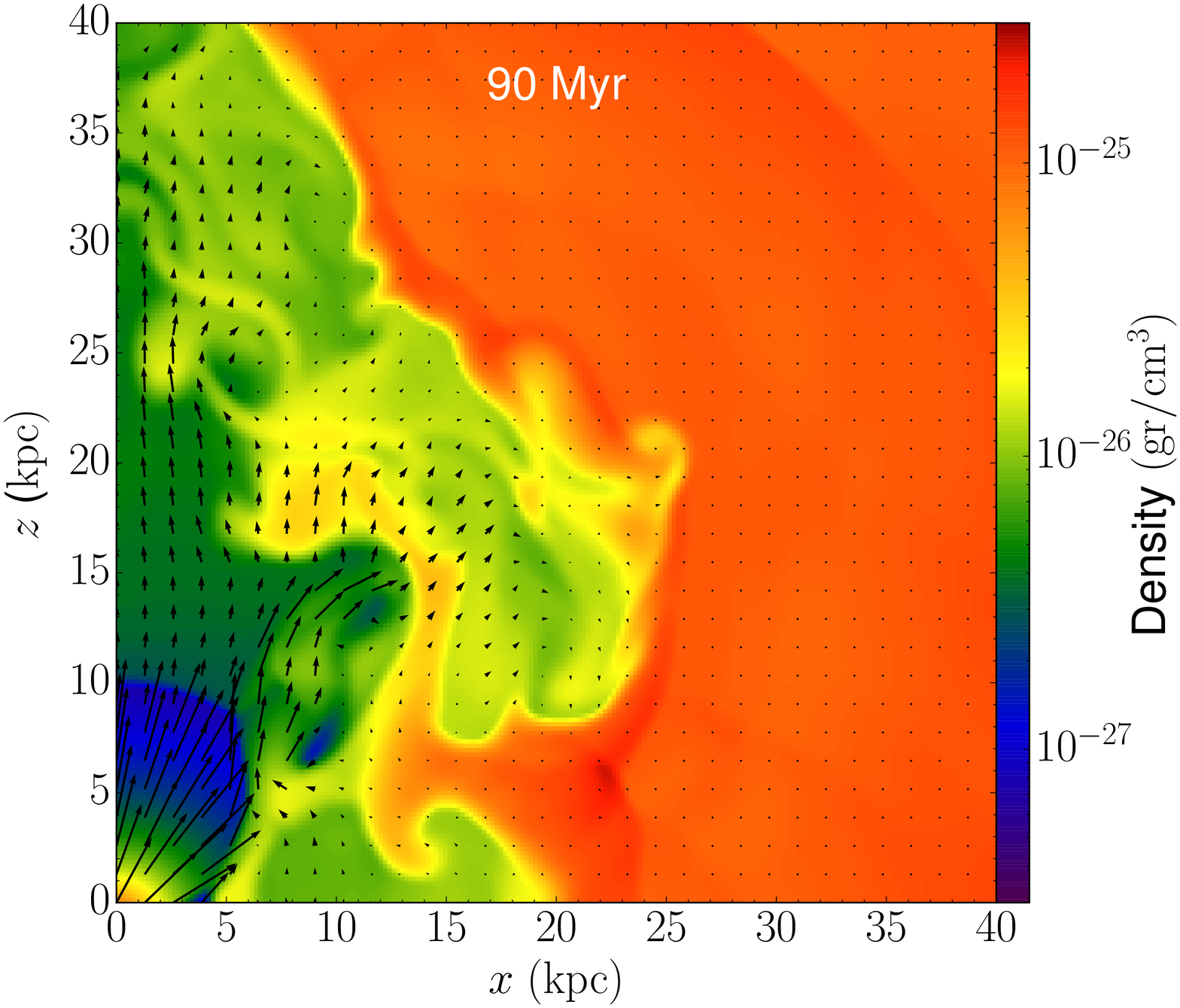}}
\hskip -0.25 cm
\subfigure{\includegraphics[width=0.24\textwidth]{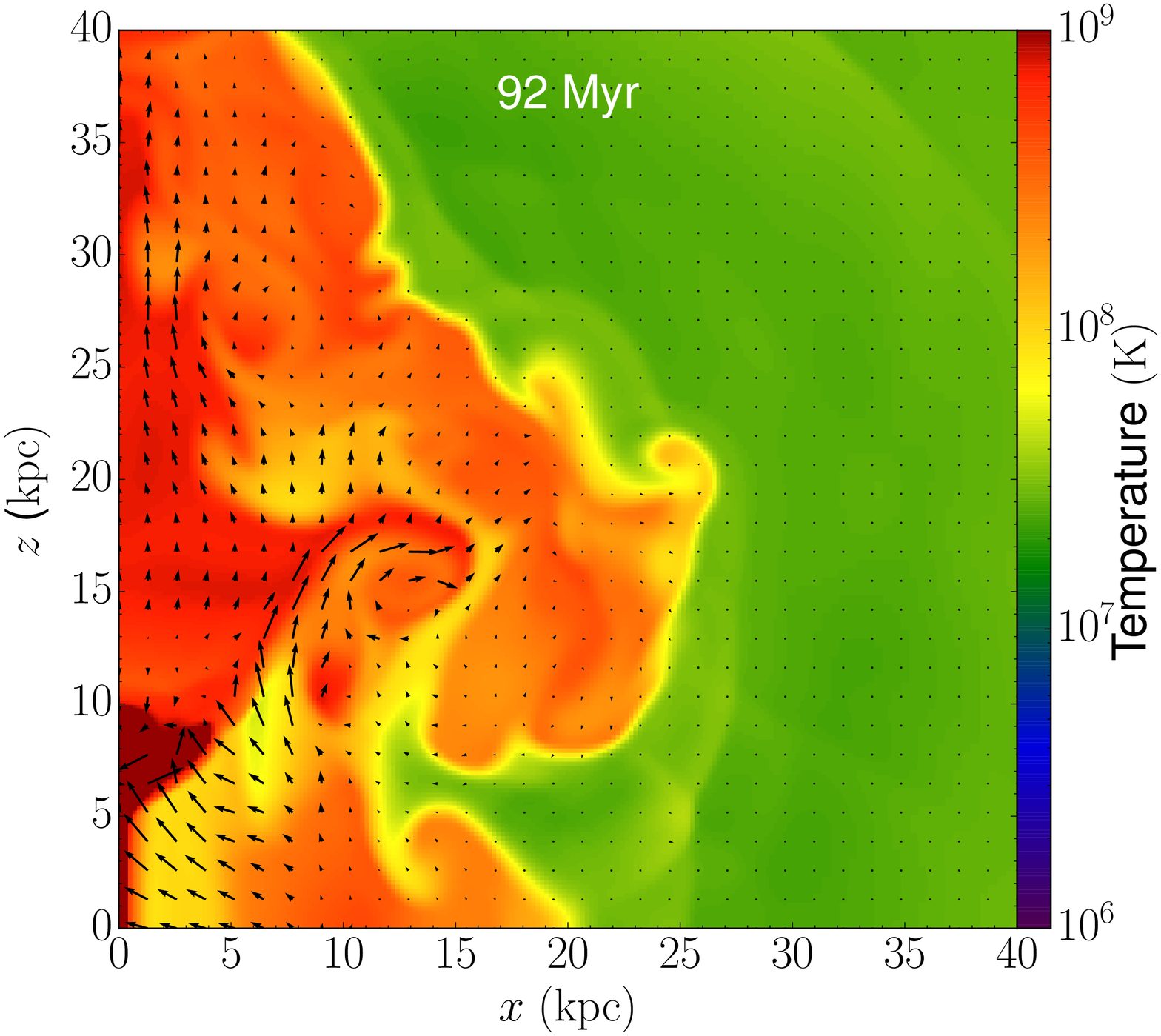}}
\subfigure{\includegraphics[width=0.24\textwidth]{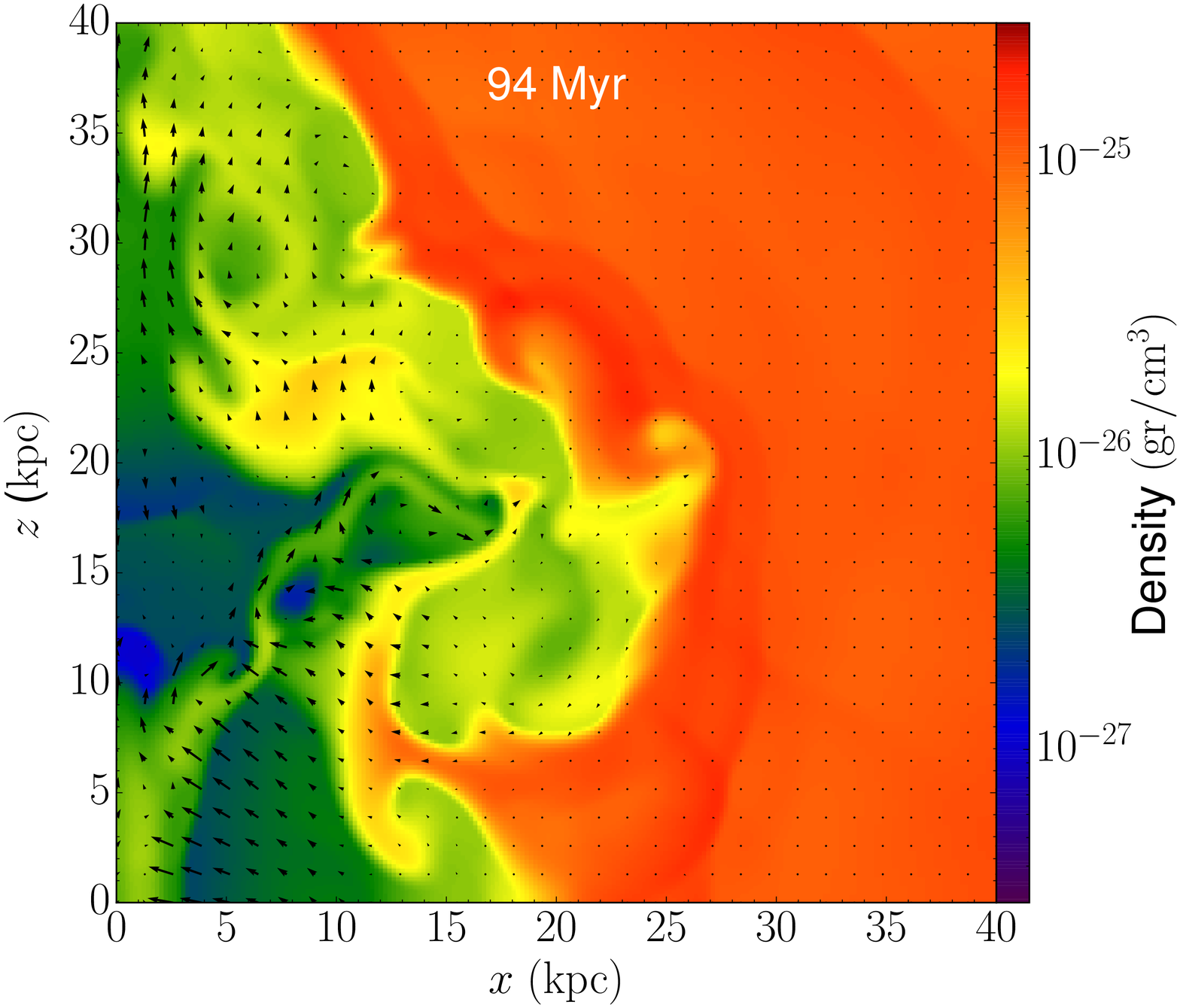}}
\hskip -0.25 cm
\subfigure{\includegraphics[width=0.24\textwidth]{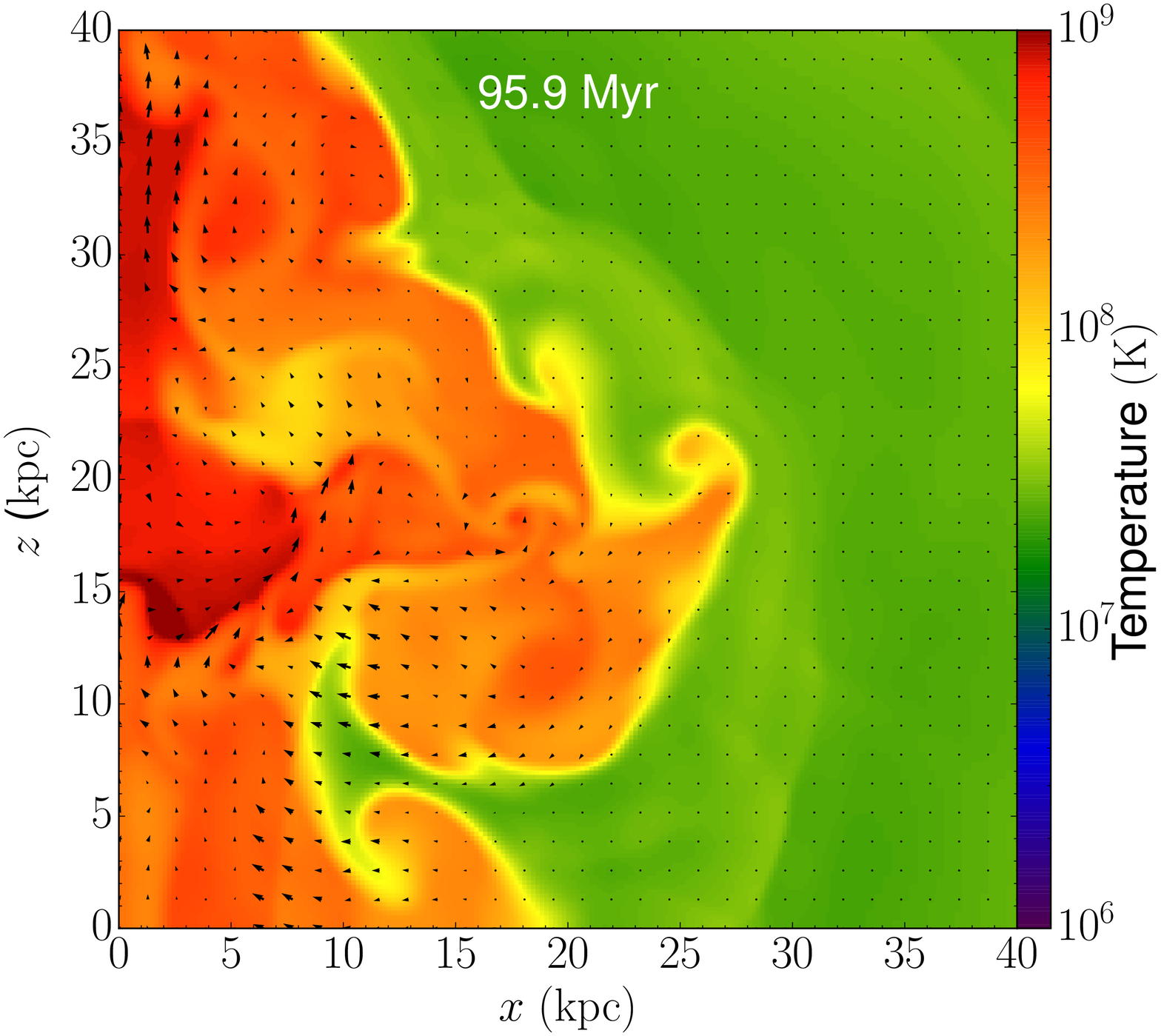}}
\subfigure{\includegraphics[width=0.24\textwidth]{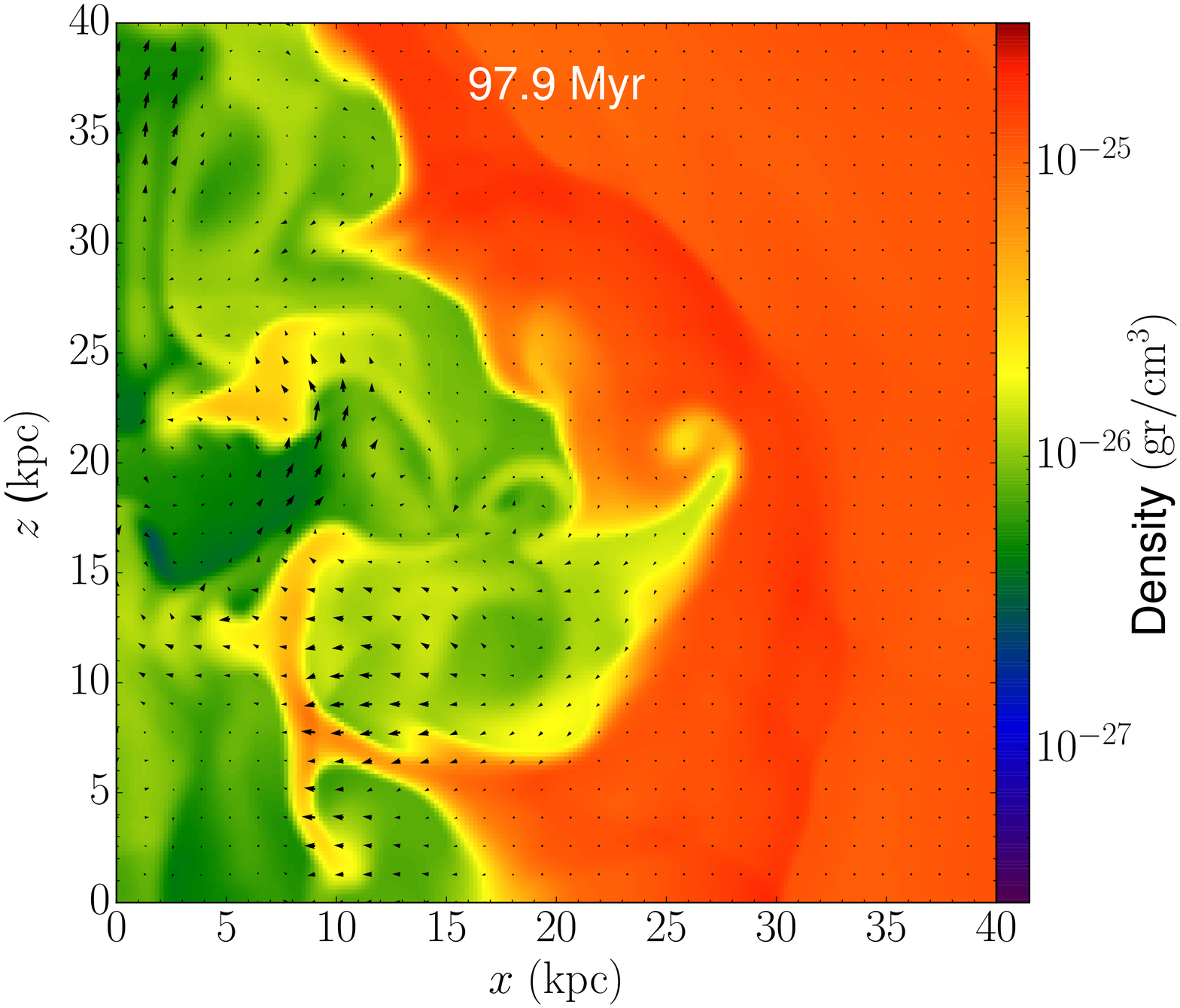}}
\hskip -0.25 cm
\subfigure{\includegraphics[width=0.24\textwidth]{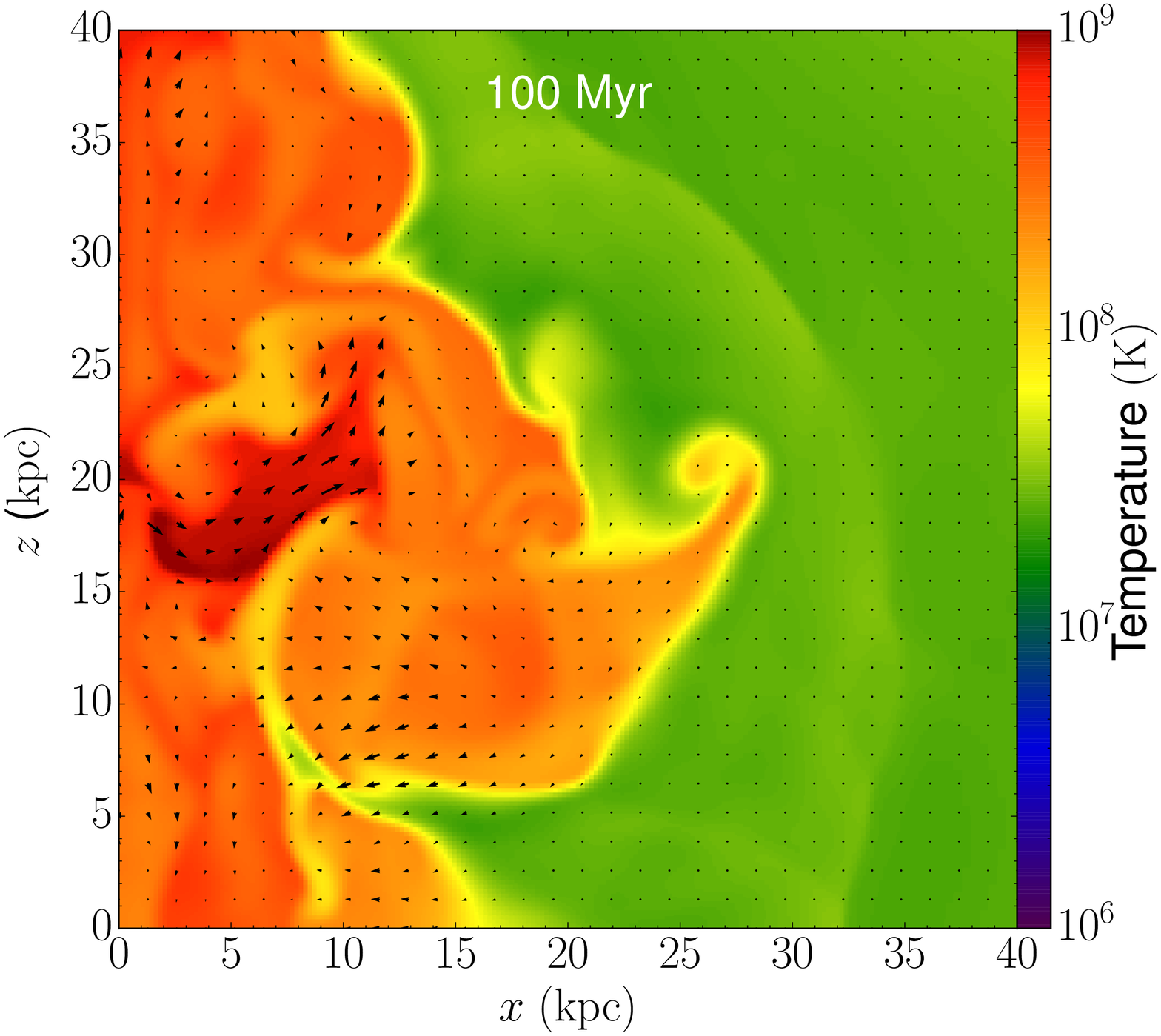}}
\caption{Density, temperature and velocity maps during the fifth activity cycle. The jet is active for the fifth time in the time period $80-90 \myr$.
Arrows show the velocity, with length proportional to the velocity magnitude. A length of $1 \kpc$ on the map corresponds to $1700 \km \s^{-1}$. When the jet is active, the length of arrows close to the origin corresponds to $8200 \kms$. }
 \label{fig:Vortices}
\end{figure}

Motivated by the new results of \cite{Hoganetal2017} and \cite{McNamaraetal2016}, in this section we show that heating by mixing operates in a gentle manner.
In Fig. \ref{fig:Vortices} we present the flow structure of the fifth activity cycle in the meridional plane $y=0$. The jet is turned on for the fifth time during the time period $80-90 \myr$, followed by a $10 \myr$ quiescent period, $90-100 \myr$. It is the quiescent period that we are interested in here.
We present the density and temperature at $t=80 \myr$, that is, $10 \myr$ after the end of the fourth active phase and at the beginning of the fifth active phase (eq. \ref{eq:jet}), followed by eight later times where we present either the density (left column) or the temperature (right column). In all panels we present the velocity map.
%FFFFFFFFFFFFFFFFFFFFFFFFFFFFFFFFFFFFFFFFFFFFFFFFFFF
%FFFFFFFFFFFFFFFFFFFFFFFFFFFFFFFFFFFFFFFFFFFFFFFFFFF

\begin{figure} % [!htb]
\centering
\centering
\subfigure{\includegraphics[width=0.24\textwidth]{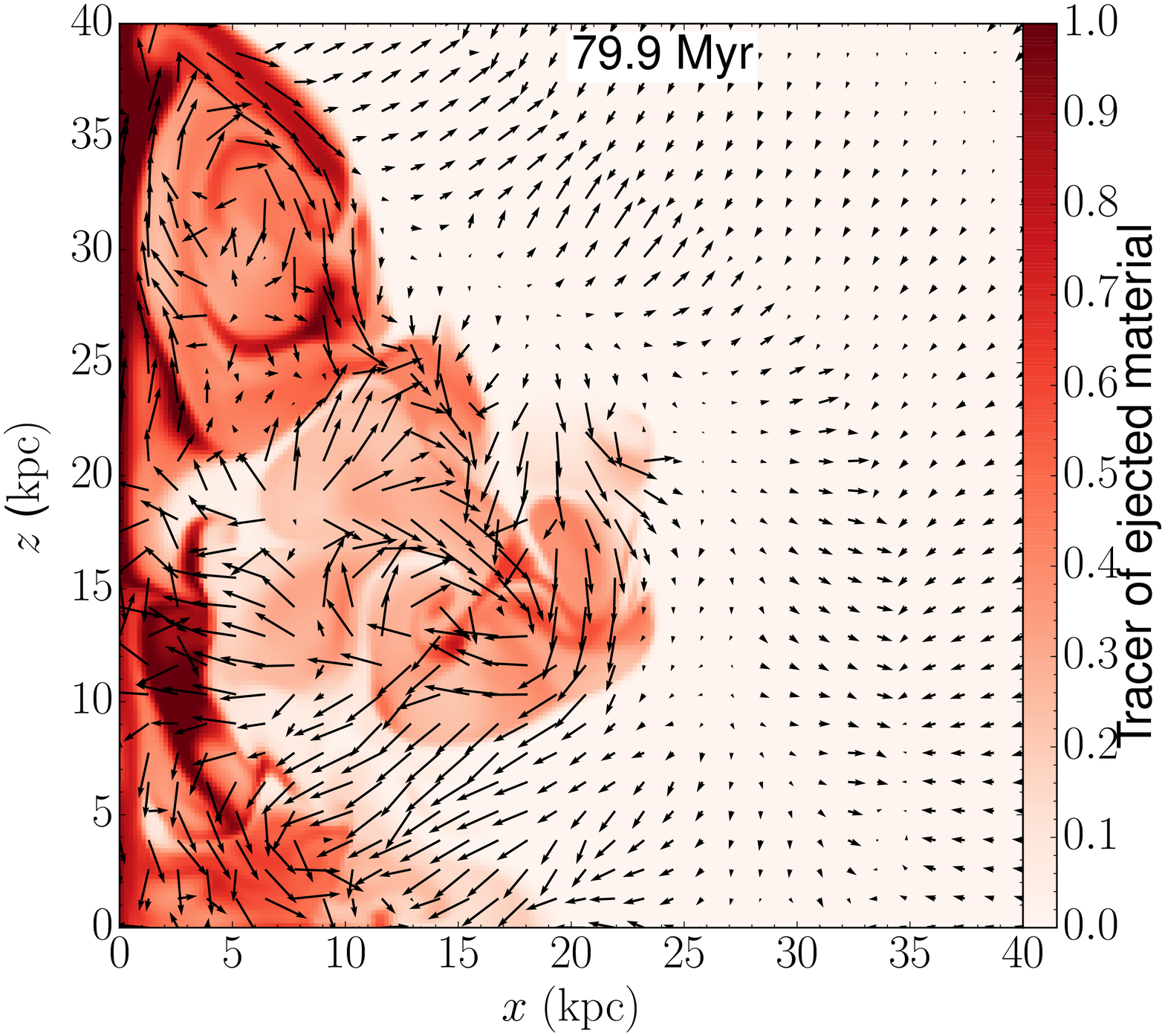}}
\hskip -0.25 cm
\subfigure{\includegraphics[width=0.24\textwidth]{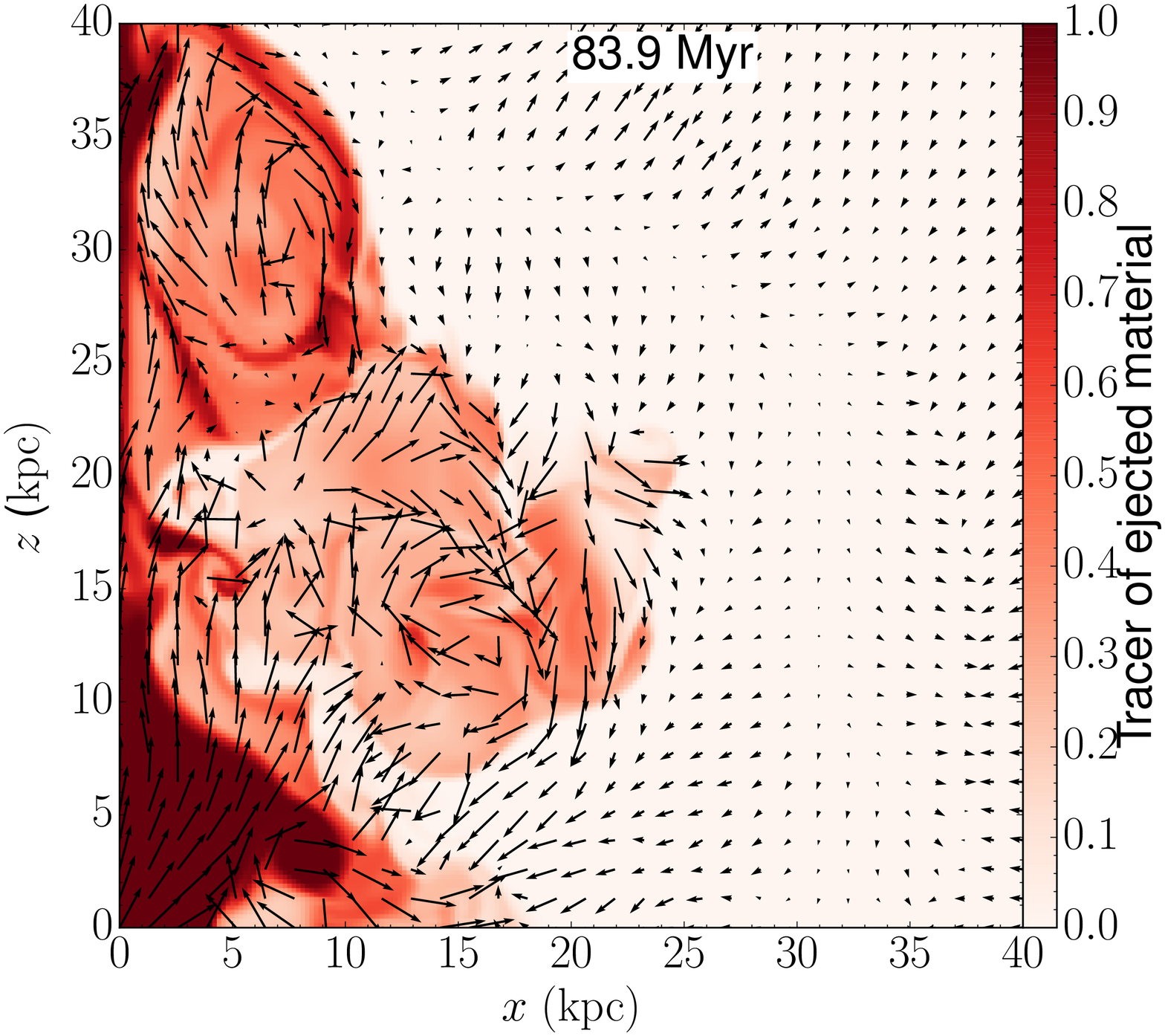}}
\subfigure{\includegraphics[width=0.24\textwidth]{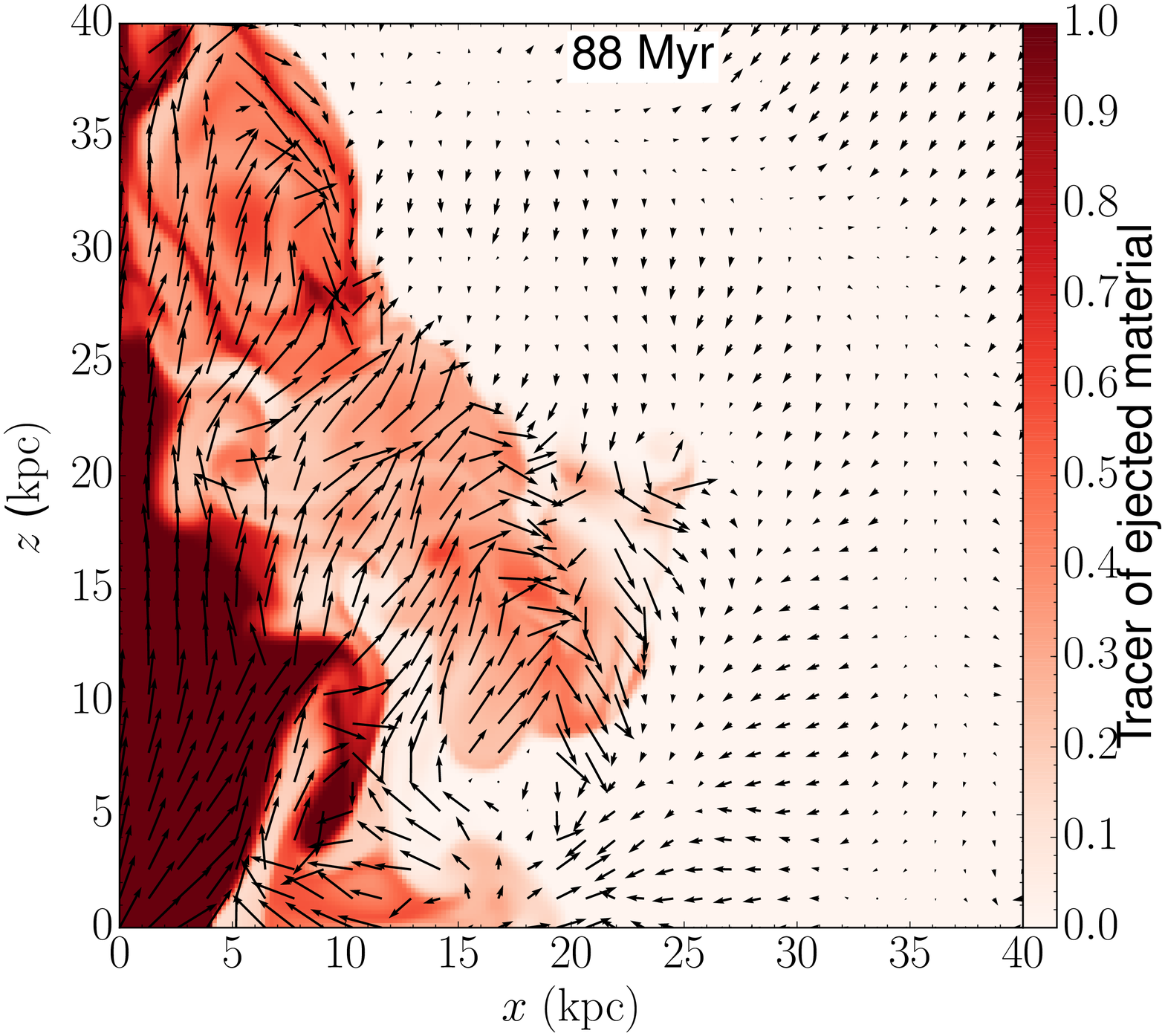}}
\hskip -0.25 cm
\subfigure{\includegraphics[width=0.24\textwidth]{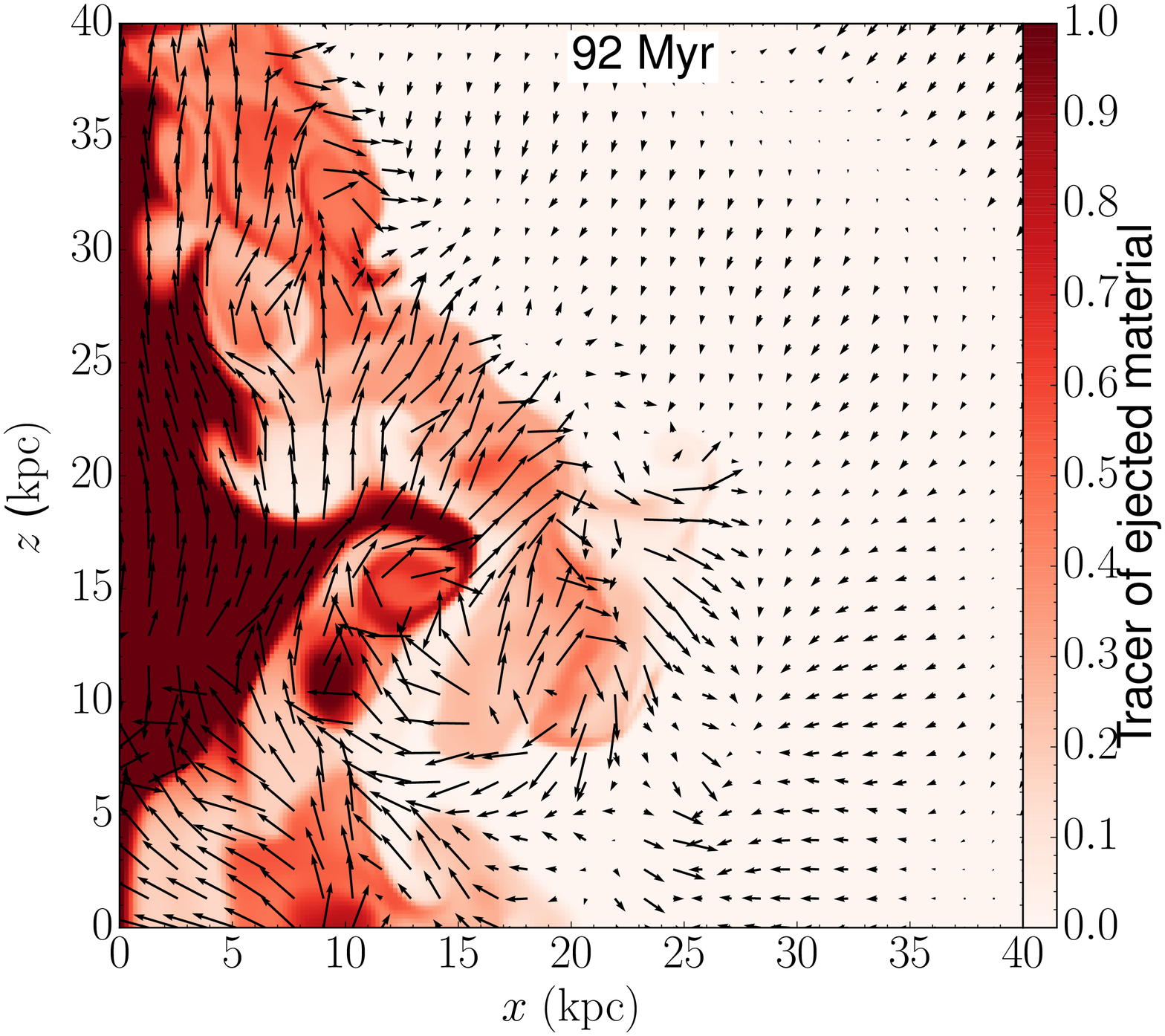}}
\subfigure{\includegraphics[width=0.24\textwidth]{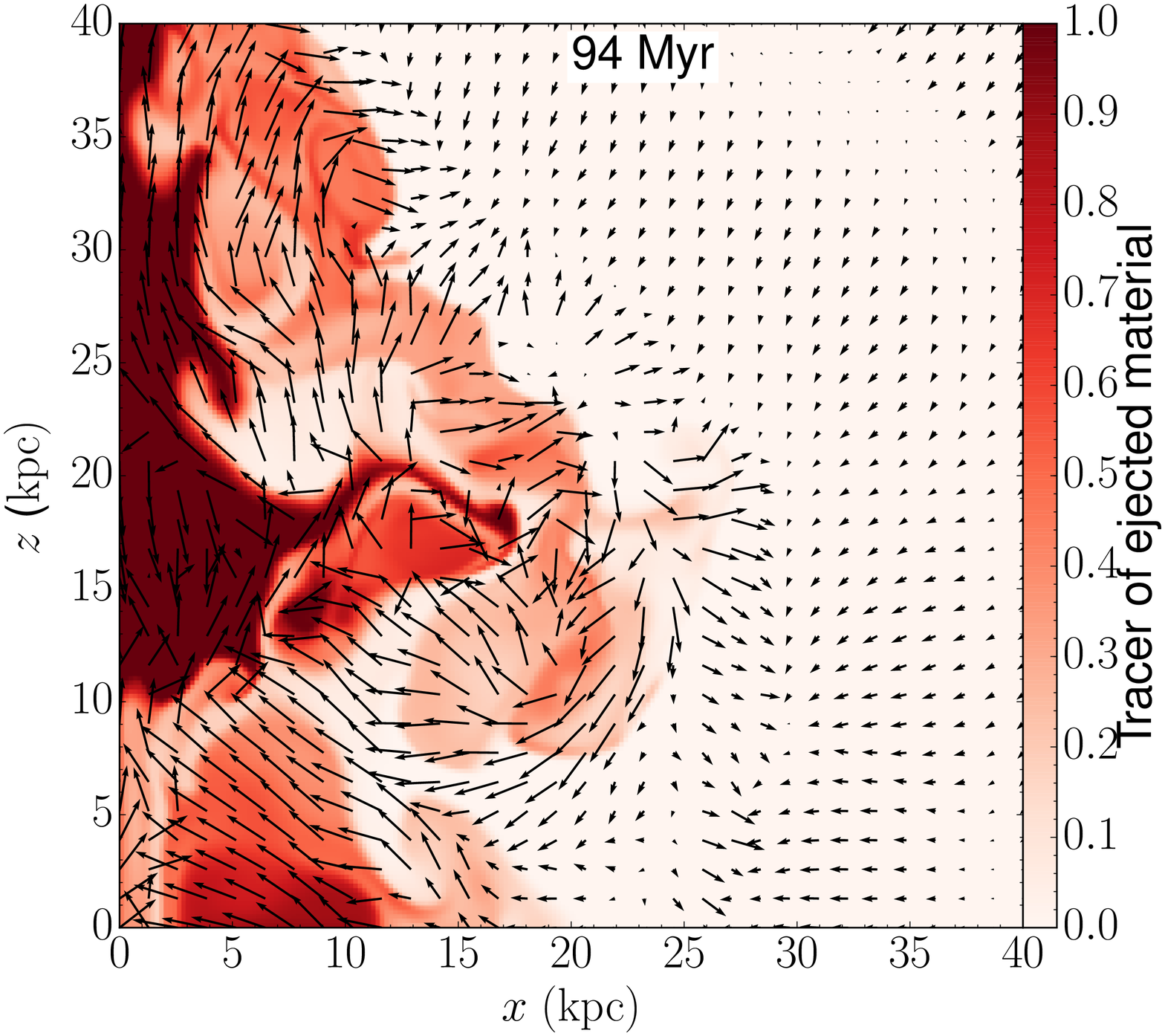}}
\hskip -0.25 cm
\subfigure{\includegraphics[width=0.24\textwidth]{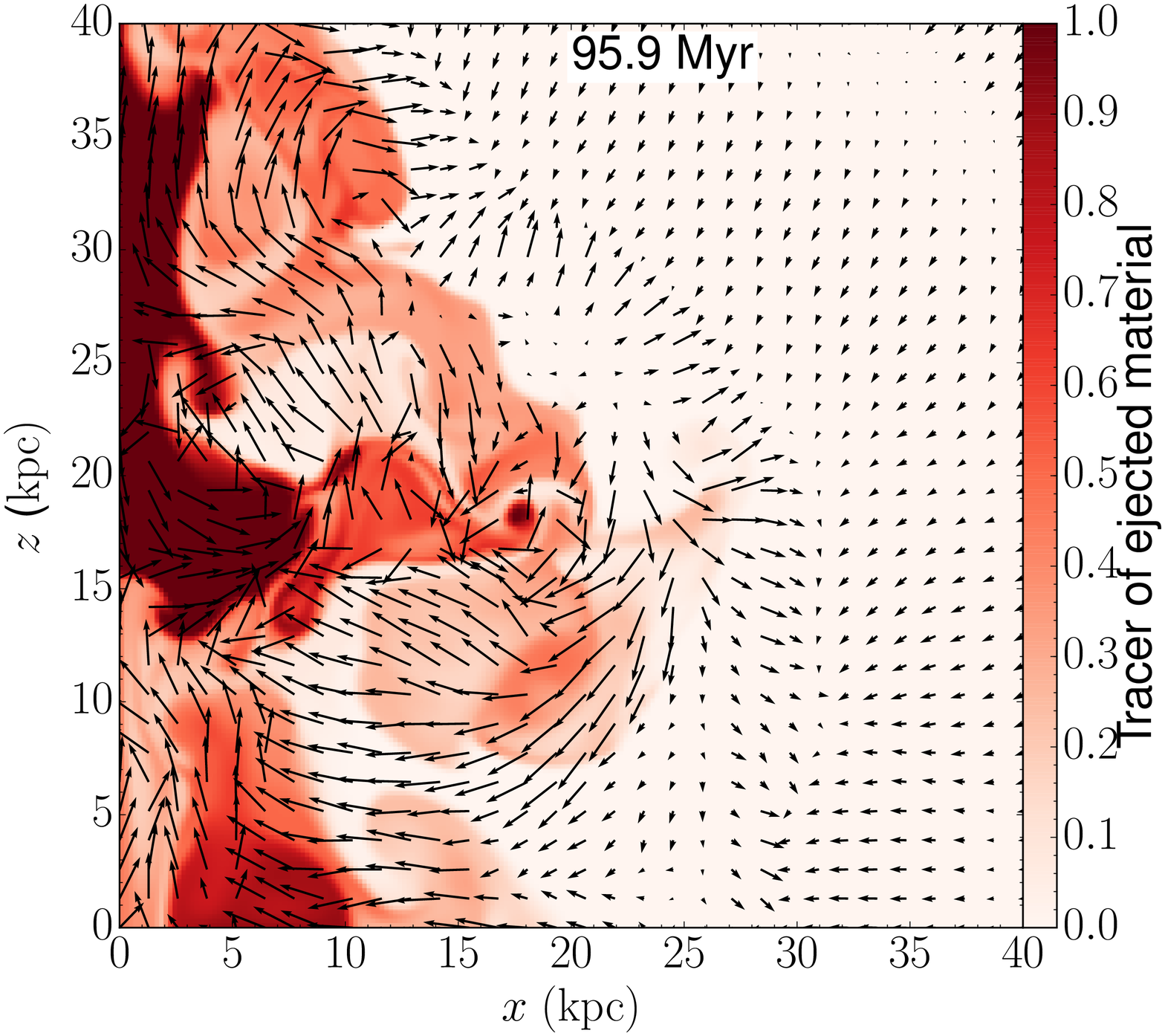}}
\subfigure{\includegraphics[width=0.24\textwidth]{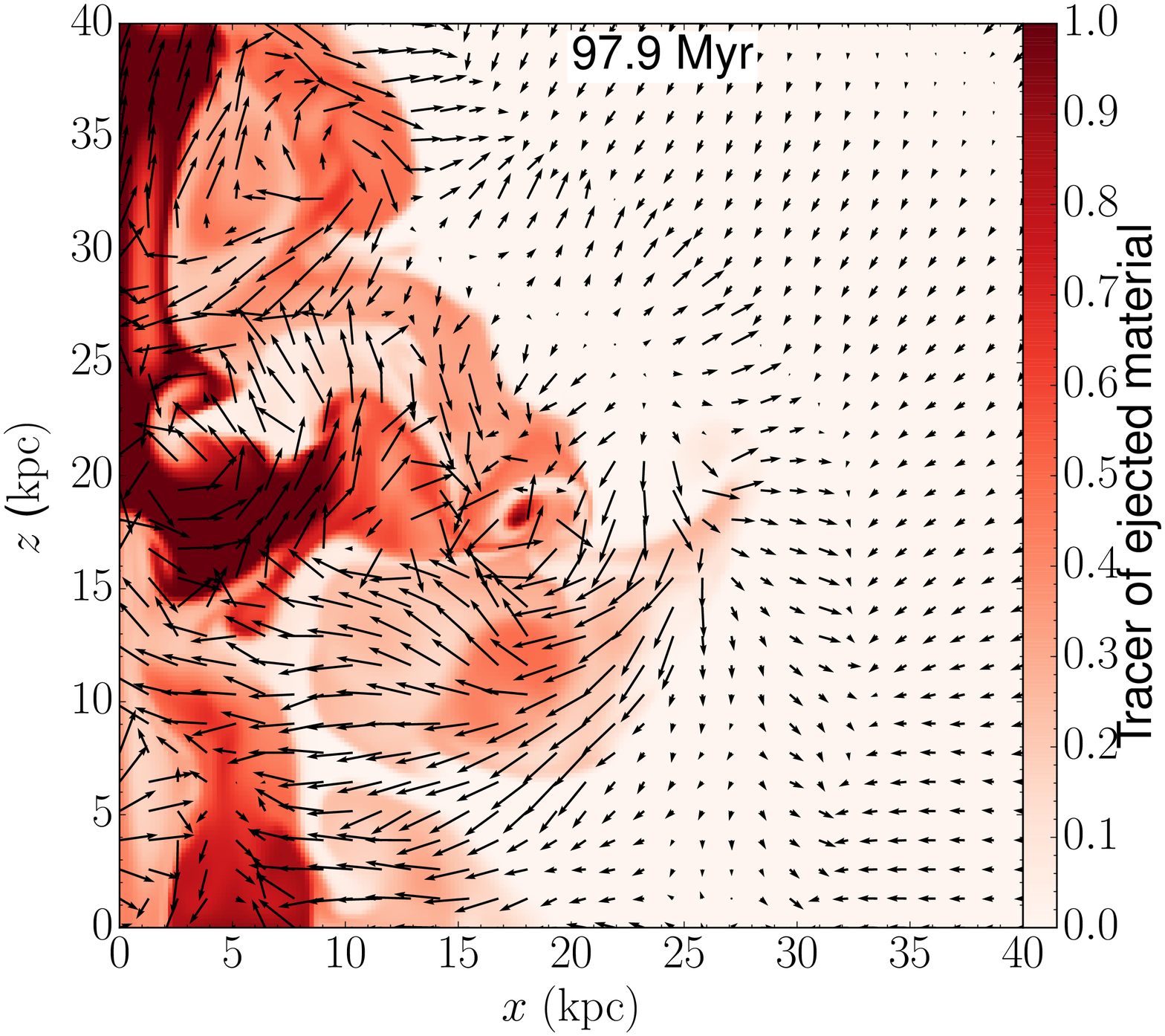}}
\hskip -0.25 cm
\subfigure{\includegraphics[width=0.24\textwidth]{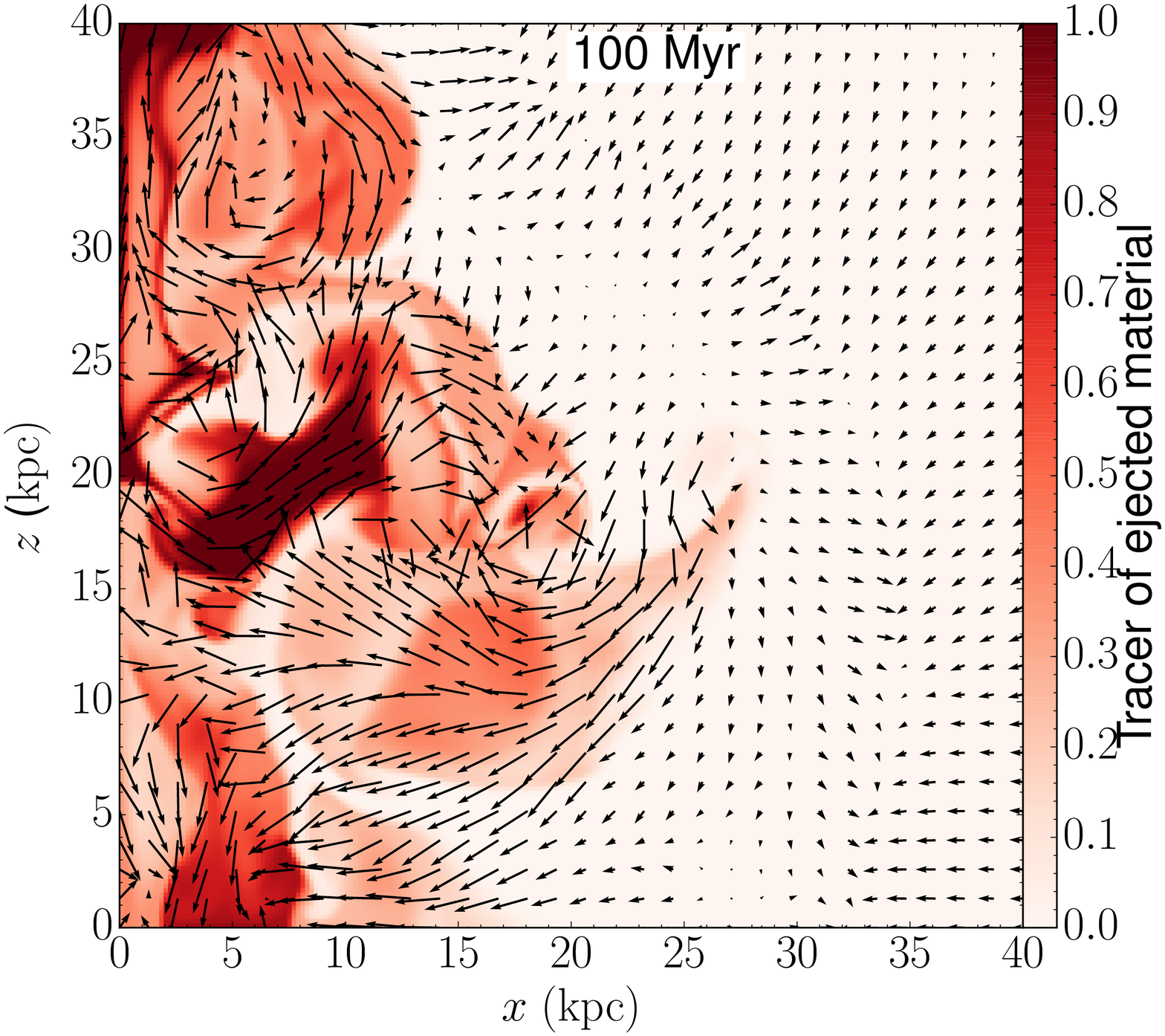}}
\caption{Evolution with time of the gas that is injected in the jet and the flow map in the meridional plane $y=0$. The color coding is for the fraction $\xi_j (t)$ at each grid point of the gas that originated in the jet.
The largest velocity vector corresponds to $v_{\rm m}=400 \kms$, a Mach number of about
$0.5$. Higher velocities are marked with arrows with the same length as that of $v_{\rm m}$. }
 \label{fig:Tracer}
\end{figure}

In addition to the velocity maps, in Fig. \ref{fig:Tracer} we follow the spreading of the gas injected in the jet by presenting the tracer of the jet's material.
The tracer is a non-physical mark that is frozen-in to the flow, and indicates the spread of the material over time. We set the initial value of the tracer of the gas that is injected into the jet to be $\xi_j(0) = 1$, and set $\xi_j (0) = 0$ for the ICM. At later times the value of $\xi_j (t)$ in each grid cell represents the fraction of the gas that started in the jet.
%FFFFFFFFFFFFFFFFFFFFFFFFFFFFFFFFFFFFFFFFFFFFFFFFFFF
%FFFFFFFFFFFFFFFFFFFFFFFFFFFFFFFFFFFFFFFFFFFFFFFFFFF

We set the velocity scale in Fig. \ref{fig:Tracer} to emphasize the flow in the ICM rather than of the post-shock jet's material, hence the velocity scale is different than in Figs, \ref{fig:EarlyFlow} and \ref{fig:Vortices}.
The vortices mix the shocked jet's material with the ICM. We take this mixing to be the main heating process of the ICM \citep{HillelSoker2016}. From the fluctuating values of the tracer of the jets $\xi_j$, even $10 \myr$ after the jet has been turned off (both at $t=80 \myr$ and $t=100 \myr$), we learn that the bubbles' gas is not fully mixed with the ICM. The vortices are still strong and the mixing process is going on. This is also seen from the non-smooth temperature maps presented in Fig. \ref{fig:Vortices}.

The conclusion from the tracer and temperature fluctuations even $10 \myr$ after the jet has been turned off, and the ongoing vortex activity, is that the heating by mixing is a continuous process that takes place on a relatively long time scale. The mixing-heating smooths out the sporadic activity of the jets launched by the AGN. This, we propose, explains the finding reported by \cite{Hoganetal2017} and \cite{McNamaraetal2016} of a gentle heating of the ICM in cooling flow clusters.

In the past we studied the mixing-heating by conducting 2D numerical simulations \citep{GilkisSoker2012, HillelSoker2014}. In those simulations each vortex is actually a torus because of the imposed axi-symmetry of the grid, and hence the results are less accurate in describing vortices. Nonetheless, we can see that the vortices live for more than $60 \myr$ after the jets have been turned off at $t=20 \myr$ (figure 6 in \citealt{GilkisSoker2012} and figures 3-5 in \citealt{HillelSoker2014}). These results of 2D simulations, although less accurate for vortices, strengthen the finding of the present analysis. Future studies should examine the behaviour of vortices during such long quiescent periods but in full 3D simulations, and for different jets' properties.

One limitation of our simulation is that it does not include physical viscosity that dissipates the kinetic energy of the vortices. We have only numerical dissipation. It is reasonable to assume that the largest vortices live for at least one revolution time $\tau_r \simeq \pi D/v$, where $D$ is the typical size of the largest vortices and $v$ is their rotational velocity. Substituting typical values for the largest vortices of $D \simeq 10 \kpc$ and $v \simeq 200 \km \s^{-1}$, we find $\tau_r \approx 10^8 \yr$. This is long enough to keep the turbulence significance even during the AGN-quiescent time periods.

% ==========================================================
\section{SUMMARY}
\label{sec:summary}
% ==========================================================

We have analysed some properties of a 3D hydrodynamical simulation of intermittent jet activity in a cooling flow cluster. This simulation was analyzed in three earlier studies \citep{HillelSoker2016, Sokeretal2016, HillelSoker2017}. In the present paper we extended the analysis and concentrated on the onset of the vortices and on the long time scale over which the vortices mix the hot bubble gas with the ICM.

{{{{ Our simulation is a generic one, i.e., it does not set to fit a specific cooling flow cluster. Although the power of the jets is at the upper end of the range of observed jet powers in cluster cores, our basic finding in a previous paper  \citep{HillelSoker2016} that heating by mixing is more important than turbulent heating and shock heating holds. The wide jets we have been using over the years are more efficient at producing vortices than narrow jets with a constant axis. However, relative motion of the jets' axis, even for narrow jets, such as a motion of the AGN relative to the ICM or precessing jets, might be more efficient at producing vortices
\citep{SternbergSoker2008precess} than wide jets with a fixed axis (as we presented here).  }}}}

In section \ref{sec:inflation} we strengthened the claim of \cite{SternbergSoker2008} that  to reveal the full properties of the jet-ICM interaction it is necessary to simulate the formation and evolution of hot bubbles by injecting propagating jets from the center of the cluster. In Fig. \ref{fig:EarlyFlow} we showed that the propagating jets form large and vigorous vortices already very close to the center. Such vortices close to the center are not formed when jets or bubbles are inserted off-center.
We think that the finding of \cite{Weinbergeretal2017} that mixing is not the main heating process of the ICM might result from that they insert jets off-center, rather than from the center.

Based on our earlier results (e.g., \citealt{HillelSoker2016}) we argue that mixing hot bubble gas with the ICM is the main heating process of the ICM. In section \ref{sec:gentle} we followed the mixing process, concentrating on the quiescent period $90-100 \myr$. From Fig. \ref{fig:Vortices} that presents fluctuations in the temperature near the center, and from Fig. \ref{fig:Tracer} that presents the fluctuations in the concentration of gas that originated in the jet, we learn that even $10 \myr$ after the jets has been turned off the mixing is not complete. We also see in these figures that the vortices still exist at that time.
These imply that the heating by mixing process operates over a long time, and it smoothes the large variations in the power of the AGN. We argued that this explains the finding reported by \cite{Hoganetal2017} and \cite{McNamaraetal2016} of a gentle heating of the ICM in cooling flow clusters.

{{{{ We thank an anonymous referee for helpful comments. }}}}
This research was supported by the Pazy Foundation.

\label{lastpage}

\begin{thebibliography}{}

\bibitem[Anderson \& Sunyaev(2016)]{AndersonSunyaev2016}  Anderson, M.,~E., \& Sunyaev, R.\ 2016, \mnras, 459, 2806

\bibitem[Arav et al.(2013)]{Aravetal2013} Arav, N., Borguet, B., Chamberlain, C., Edmonds, D., \& Danforth, C.\ 2013, \mnras, 436, 3286

\bibitem[Ar{\'e}valo et al.(2016)]{Arevalo2016} Ar{\'e}valo, P., Churazov, E., Zhuravleva, I., Forman, W.~R., \& Jones, C.\  2016, \apj, 818, 14

\bibitem[Banerjee \& Sharma(2014)]{BanerjeeSharma2014} Banerjee, N., \& Sharma, P.\ 2014, \mnras, 443, 687

\bibitem[Barai et al.(2016)]{Baraietal2016} Barai, P., Murante, G., Borgani, S., Gaspari, Massimo., Granato, G.~L., Monaco, P., \& Ragone-Figueroa. Cinthia\ 2016, \mnras, 461, 1548

\bibitem[Br{\"u}ggen \& Kaiser(2002)]{BruggenKaiser2002} Br{\"u}ggen, M., \& Kaiser, C.~R.\ 2002, \nat, 418, 301

\bibitem[Br{\"u}ggen et al.(2009)]{Bruggenetal2009} Br{\"u}ggen, M., Scannapieco, E., \& Heinz, S.\ 2009, \mnras, 395, 2210

\bibitem[Choudhury \& Sharma(2016)]{ChoudhurySharma2016} Choudhury, P.~P., \& Sharma, P.\ 2016, \mnras, 457, 2554

\bibitem[De Young(2010)]{DeYoung2010} De Young, D.~S.\ 2010, \apj, 710, 743

\bibitem[Donahue et al.(2017)]{Donahueetal2017} Donahue, M., Connor, T., Voit, G.~M., \& Postman, M.\ 2017, \apj, 835, 216

\bibitem[Fabian(2012)]{Fabian2012} Fabian, A.~C.\ 2012, \araa, 50, 455

\bibitem[Fabian et al.(2006)]{Fabianetal2006} Fabian, A.~C., Sanders, J.~S., Taylor, G.~B., Allen, S.~W., Crawford, C.~S., Johnstone, R.~M., \& Iwasawa, K.\ 2006, \mnras, 366, 417

\bibitem[Fabian et al.(2017)]{Fabianetal2017} Fabian, A.~C., Walker, S.~A., Russell, H.~R.,  Pinto, C., Sanders, J.~S., \& Reynolds, C.~S.\ 2017, \mnras, 464, L1

\bibitem[Falceta-Gon{\c c}alves et al.(2010)]{Falcetaetal2010} Falceta-Gon{\c c}alves, D., de Gouveia Dal Pino, E.~M., Gallagher, J.~S., \& Lazarian, A.\ 2010, \apjl, 708, L57

\bibitem[Farage et al.(2012)]{Farage2012} Farage, C.~L., McGregor, P.~J., \& Dopita, M.~A.\ 2012, \apj, 747, 28

\bibitem[Forman et al.(2007)]{Formanetal2007}  Forman, W., Jones, C., Churazov, E., et al.\ 2007, \apj, 665, 1057

\bibitem[Fujita et al.(2013)]{Fujitaetal2013} Fujita, Y., Kimura, S., \& Ohira, Y.\ 2013, \mnras, 432, 1434

\bibitem[Fujita \& Ohira(2013)]{FujitaOhira2013} Fujita, Y., \& Ohira, Y.\ 2013, \mnras, 428, 599

\bibitem[Gaspari(2015)]{Gaspari2015} Gaspari, M.\ 2015, \mnras, 451, L60

\bibitem[Gaspari et al.(2013)]{Gasparietal2013} Gaspari, M., Brighenti, F., \& Ruszkowski, M.\ 2013, Astronomische Nachrichten, 334, 394

\bibitem[Gaspari et al.(2014)]{Gasparietal2014} Gaspari, M., Churazov, E., Nagai, D., Lau, E.~T., \& Zhuravleva, I.\ 2014, \aap, 569, A67

\bibitem[Gaspari \& S{\c a}dowski(2017)]{GaspariSadowski2017} Gaspari, M., \& S{\c a}dowski, A.\ 2017, \apj, 837, 149

\bibitem[Gaspari et al.(2017)]{Gasparietal2017} Gaspari, M., Temi, P., \& Brighenti, F.\ 2017, \mnras, 466, 677

\bibitem[Gilkis \& Soker(2012)]{GilkisSoker2012} Gilkis, A., \& Soker, N.\ 2012, \mnras, 427, 1482

\bibitem[Guo \& Oh(2008)]{GuoOh2008} Guo, F., \& Oh, S.~P.\ 2008, \mnras, 384, 251

\bibitem[Hamer et al.(2016)]{Hameretal2016} Hamer, S.~L., Edge, A.~C., Swinbank, A.~M., et al.\ 2016, \mnras, 460, 1758

\bibitem[Hillel \& Soker(2014)]{HillelSoker2014} Hillel, S., \& Soker, N.\ 2014, \mnras, 445, 4161

\bibitem[Hillel \& Soker(2016)]{HillelSoker2016} Hillel, S., \& Soker, N.\ 2016, \mnras, 455, 2139

\bibitem[Hillel \& Soker(2017)]{HillelSoker2017} Hillel, S., \& Soker, N.\ 2017, \mnras, 466, L39

\bibitem[Hitomi Collaboration et al.(2016)]{Hitomi2016} Hitomi Collaboration\ 2016, \nat, 535, 117

\bibitem[Hofmann et al.(2016)]{Hofmannetal2016} Hofmann, F., Sanders, J.~S., Nandra, K., Clerc, N., \& Gaspari, M.\ 2016, \aap, 585, A130

\bibitem[Hogan et al.(2017)]{Hoganetal2017} Hogan, M.~T., McNamara, B.~R., Pulido, F., et al.\ 2017, arXiv:1704.00011

\bibitem[Li et al.(2015)]{Lietal2015} Li, Y., Bryan, G.~L., Ruszkowski, M., Voit, G.~M., O'Shea, B.~W., \& Donahue, M.\ 2015, \apj, 811, 73

\bibitem[Loubser et al.(2016)]{Loubseretal2016} Loubser, S.~I., Babul, A., Hoekstra, H., Mahdavi, A., Donahue, M., Bildfell, C., \& Voit, G.~M.\ 2016, \mnras, 456, 1565

\bibitem[McNamara \& Nulsen(2012)]{McNamaraNulsen2012} McNamara, B.~R., \& Nulsen, P.~E.~J.\ 2012, New Journal of Physics, 14, 055023

\bibitem[McNamara et al.(2016)]{McNamaraetal2016} McNamara, B.~R., Russell, H.~R., Nulsen, P.~E.~J.,  Hogan, M.~T., Fabian, A.~C., Pulido, F., \& Edge, A.~C.\ 2016, \apj, 830, 79

\bibitem[Meece et al.(2017)]{Meeceetal2017} Meece, G.~R., Voit, G.~M., \& O'Shea, B.~W.\ 2017, \apj, 841, 133

\bibitem[Mignone et al.(2007)]{Mignone2007} Mignone, A., Bodo, G., Massaglia, S., et al.\ 2007, \apjs, 170, 228

\bibitem[Pfrommer(2013)]{Pfrommer2013} Pfrommer, C.\ 2013, \apj, 779, 10

\bibitem[Pizzolato \& Soker(2005)]{PizzolatoSoker2005} Pizzolato, F., \& Soker, N.\ 2005, \apj, 632, 821

\bibitem[Prasad et al.(2015)]{Prasadetal2015} Prasad, D., Sharma, P., \& Babul, A.\ 2015, \apj, 811, 108

\bibitem[Prasad et al.(2016)]{Prasadetal2016} Prasad, D., Sharma, P., \& Babul, A.\ 2016, arXiv:1611.02710

\bibitem[Randall et al.(2015)]{Randalletal2015} Randall, S.~W., Nulsen,
P.~E.~J., Jones, C., et al.\ 2015, \apj, 805, 112

\bibitem[Reynolds et al.(2015)]{Reynoldsetal2015} Reynolds, C.~S., Balbus, S.~A., \& Schekochihin, A.~A.\ 2015, \apj, 815, 41

\bibitem[Russell et al.(2016)]{Russelletal2016} Russell, H.~R., McNamara, B.~R., Fabian, A.~C., et al.\ 2016, \mnras, 458, 3134

\bibitem[Singh \& Sharma(2015)]{SinghSharma2015} Singh, A., \& Sharma, P.\ 2015, \mnras, 446, 1895
%\bibitem[Soker(2006)]{Soker2006} Soker, N.\ 2006, New Astronomy, 12, 38

\bibitem[Soker(2016)]{Soker2016} Soker, N.\ 2016, New Astronomy Reviews, 75, 1

\bibitem[Soker et al.(2013)]{Sokeretal2013} Soker, N., Akashi, M., Gilkis, A., Hillel, S., Papish, O., Refaelovich, M., \& Tsebrenko, D.\ 2013, Astronomische Nachrichten, 334, 402

\bibitem[Soker et al.(2016)]{Sokeretal2016} Soker, N., Hillel, S., \& Sternberg, A.\ 2016, Research in Astronomy and Astrophysics, 16, 015

\bibitem[Sternberg \& Soker(2008a)]{SternbergSoker2008precess} Sternberg, A., \& Soker, N.\ 2008a, \mnras, 384, 1327

\bibitem[Sternberg \& Soker(2008b)]{SternbergSoker2008} Sternberg, A., \& Soker, N.\ 2008b, \mnras, 389, L13

\bibitem[Sternberg \& Soker(2009)]{SternbergSoker2009} Sternberg, A., \& Soker, N.\ 2009, \mnras, 395, 228

\bibitem[Sutherland \& Dopita(1993)]{SutherlandDopita1993} Sutherland, R.~S., \& Dopita, M.~A.\ 1993, \apjs, 88, 253

\bibitem[Tremblay et al.(2015)]{Tremblayetal2015} Tremblay, G.~R., O'Dea, C.~P., Baum, S.~A., et al.\ 2015, \mnras, 451, 3768]

\bibitem[Tremblay et al.(2016)]{Tremblayetal2016} Tremblay, G.~R., Oonk, J.~B.~R., Combes, F., et al.\ 2016, \nat,  534, 218

\bibitem[Valentini \& Brighenti(2015)]{ValentiniBrighenti2015} Valentini, M., \& Brighenti, F.\ 2015, \mnras, 448, 1979

\bibitem[Vernaleo \& Reynolds(2006)]{VernaleoReynolds2006} Vernaleo, J.~C., \& Reynolds, C.~S.\ 2006, \apj, 645, 83

\bibitem[Voit \& Donahue(2015)]{VoitDonahue2015} Voit, G.~M., \& Donahue, M.\ 2015, \apjl, 799, L1

\bibitem[Voit et al.(2015)]{Voitetal2015} Voit, G.~M., Donahue, M., Bryan, G.~L., \& McDonald, M.\ 2015, \nat, 519, 203

\bibitem[Voit et al.(2017)]{Voitetal2017} Voit, G.~M., Meece, G., Li, Y., O'Shea, B.~W., Bryan, G.~L., \& Donahue, M.\ 2017, arXiv:1607.02212

\bibitem[Weinberger et al.(2017)]{Weinbergeretal2017} Weinberger, R., Ehlert, K., Pfrommer, C., Pakmor, R., \& Springel, V.\ 2017, arXiv:1703.09223

\bibitem[Yang \& Reynolds(2016)]{YangReynolds2016b} Yang, H.-Y.~K., \& Reynolds, C.~S.\ 2016, \apj, 829, 90

\bibitem[Zhuravleva et al.(2017)]{Zhuravlevaetal2017} {{{{ Zhuravleva, I., Allen, S.~W., Mantz, A.~B., \& Werner, N.\ 2017, arXiv:1707.02304 }}} }

\bibitem[Zhuravleva et al.(2015)]{Zhuravlevaetal2015} Zhuravleva, I.,
Churazov, E., Arevalo, P., et al.\ 2015,  \mnras, 450, 4184

\bibitem[Zhuravleva et al.(2014)]{Zhuravlevaetal2014} Zhuravleva, I.,
Churazov, E., Schekochihin, A.~A., et al.\ 2014, \nat, 515, 85

\end{thebibliography}
\end{document}